\documentclass[%
 aip,
 jmp,
 amsmath,amssymb,
 reprint,%
]{revtex4-1}

\usepackage{graphicx}
\usepackage{dcolumn}
\usepackage{bm}

\usepackage{amsfonts,mathtools}
\usepackage[colorlinks=true]{hyperref}

\usepackage{physics} 

\usepackage{xcolor}  

\usepackage{easybmat}

\usepackage{tikz}
\usetikzlibrary{shapes}

\usepackage[figurename=Fig.]{caption}
\usepackage[font={footnotesize}]{subcaption}

\newcommand{\si}[1]{#1}

\DeclareMathOperator{\dal}{\Box}  
\DeclareMathOperator{\Curl}{\mathrm{curl}}   
\DeclareMathOperator{\Div}{\mathrm{div}}

\newcommand{\Ort}[1][3]{\mathrm{O}(#1)}
\newcommand{\SO}[1][3]{\mathrm{SO}(#1)}
\newcommand{\LG}[1][3]{\mathrm{O}(#1,1)} 
\newcommand{\LR}[1][3]{\smash{\mathrm{SO}(#1,1)^\uparrow}}

\newcommand{\gaussel}{\raisebox{2pt}{\tikz{\draw[very thick, orange!70!black!50](0,0) -- (5mm,0);}}}

\newcommand{\gaussmag}{\raisebox{2pt}{\tikz{\draw[very thick,cyan!50!black!60](0,0) -- (5mm,0);}}}

\newcommand{\ampere}{\raisebox{2pt}{\tikz{\draw[ very thick,purple!30!black!40](0,0) -- (5mm,0);}}}

\newcommand{\faraday}{\raisebox{2pt}{\tikz{\draw[ very thick,green!30!black!50](0,0) -- (5mm,0);}}}

\newcommand{\electric}{\raisebox{0pt}{\tikz{\node[scale=0.6,circle,thick,draw=red!50!black!,fill=red!20!,fill opacity=0.5](){};}}}

\newcommand{\magnetic}{\raisebox{0pt}{\tikz{\node[scale=0.6,circle,thick,draw=blue!50!black!,fill=blue!20!,fill opacity=0.5](){};}}}

\newcommand{\charge}{\raisebox{0pt}{\tikz{\node[scale=0.6,thick,regular polygon, regular polygon sides=4,draw=orange!80!black!60,fill=orange!80!black!15](){};}}}

\newcommand{\current}{\raisebox{0pt}{\tikz{\node[scale=0.6,thick,regular polygon, regular polygon sides=4,draw=purple!70!black!,fill=purple!70!black!15](){};}}}

\makeatletter
\def\@email#1#2{%
 \endgroup
 \patchcmd{\titleblock@produce}
  {\frontmatter@RRAPformat}
  {\frontmatter@RRAPformat{\produce@RRAP{*#1\href{mailto:#2}{#2}}}\frontmatter@RRAPformat}
  {}{}
}%
\makeatother
\begin{document}

\preprint{AIP/123-QED}

\title[Dimensional reduction of electromagnetism]{Dimensional reduction of electromagnetism}
\author{Rocco Maggi}
 \affiliation{Dipartimento Interateneo di Fisica, Universit\`a di Bari, I-70126 Bari, Italy}
 \affiliation{Istituto Nazionale di Fisica Nucleare, Sezione di Bari, I-70125 Bari, Italy}
 
\author{Elisa Ercolessi}
 \affiliation{Dipartimento di Fisica e Astronomia, Universit\`a di Bologna, I-40127 Bologna, Italy}
 \affiliation{Istituto Nazionale di Fisica Nucleare, Sezione di Bologna, I-40127 Bologna, Italy}

\author{Paolo Facchi}
 \affiliation{Dipartimento Interateneo di Fisica, Universit\`a di Bari, I-70126 Bari, Italy}
 \affiliation{Istituto Nazionale di Fisica Nucleare, Sezione di Bari, I-70125 Bari, Italy}
 
\author{Giuseppe Marmo}
 \affiliation{Dipartimento di Fisica ``E. Pancini'', Universit\`a di Napoli Federico II, I-80126, Naples, Italy}
 \affiliation{Istituto Nazionale di Fisica Nucleare, Sezione di Napoli, I-80126, Naples, Italy}
 
\author{\\Saverio Pascazio}
 \affiliation{Dipartimento Interateneo di Fisica, Universit\`a di Bari, I-70126 Bari, Italy}
 \affiliation{Istituto Nazionale di Fisica Nucleare, Sezione di Bari, I-70125 Bari, Italy}
 
\author{Francesco V. Pepe}
 \affiliation{Dipartimento Interateneo di Fisica, Universit\`a di Bari, I-70126 Bari, Italy}
 \affiliation{Istituto Nazionale di Fisica Nucleare, Sezione di Bari, I-70125 Bari, Italy}


\date{\today}

\begin{abstract}
We derive one- and two-dimensional models for classical electromagnetism by making use of Hadamard's method of descent. Low-dimensional electromagnetism is conceived as a specialization of the higher dimensional one, in which the fields are uniform along the additional spatial directions. This strategy yields two independent electromagnetisms in two spatial coordinates, and four independent electromagnetisms in one spatial coordinate.  
\end{abstract}

\maketitle

\section{Introduction}
Questions about space, its flatness, and its number of dimensions have always fascinated physicists~\cite{Ehrenfest}, mathematicians~\cite{hardy}, and novelists alike~\cite{alice,petitprince,flatland}. We live in three spatial dimensions, but often wonder about worlds with a different number of dimensions, and try to build a consistent picture of such worlds, whose properties are beyond our sensorial perceptions, but certainly not out of the reach of our imagination. 

Ehrenfest, one century ago, wondered whether the fundamental laws of physics ``require'', in some sense, that there be three (space) dimensions~\cite{Ehrenfest}. 
Ehrenfest realized that there are a number of geometric peculiarities that make three dimensions different.
He observed that (i) in a generic $(n+1)$-dimensional space-time the number of components of the electric field is equal to the number of independent boosts, which in turn is equal to the number $n$ of spatial coordinates; and (ii) the number of components of the magnetic field is equal to the number of independent rotations $\binom{n}{2}=n(n-1)/2$, coinciding with the number of distinct two-dimensional planes; (iii) these two numbers are equal to each other, and to $3$, only for $n=3$, while, for all other values of $n$, the electric and magnetic fields are objects of different kinds. From this perspective, the laws of electromagnetism entail the three-dimensionality of space because the ``dualism between the electric and magnetic quantities'' is a characteristic of our world.
According to Ehrenfest, three (and four) dimensions appear different and somehow more interesting. These seminal intuitions have been corroborated by mathematical and physical research in the decades that followed. It is curious that such technical observations motivate scientists to ask bold questions, that even philosophers would not dare to formulate. One such question is: how would the world look like in \emph{two}, or even \emph{one} space dimension?

In this article we will not tackle this general and very difficult question, and will rather content ourselves with a simpler query. We will ask which form can the laws of electromagnetism (EM) take in $n<3$ space dimensions. A moment's reflection shows that even this humbler question is void of meaning, unless one specifies a ``procedure'' that enables one to reduce the number of dimensions. In particular, what should one demand to a one or two-dimensional world? Which symmetries, fields, potentials, Lorentz force, Maxwell's equations, vector products, relativistic and gauge invariance should apply? Is it possible to formulate EM in $n<3$ space dimensions in such a way that \emph{all} the above properties are consistently defined? 

In order to formulate EM in $n<3$ space dimensions, we will adopt a strategy proposed by Hadamard~\cite{Hadamard} one century ago, known as \emph{method of descent}. The idea, in Hadamard's words, ``consists in noticing that he who can do more can do less: if we can integrate equations with $m$ variables, we can do the same for equations with $m-1$ variables''.
Hadamard's primary objective was to solve a differential equation, such as the wave and heat equations, depending on a set of $n$ independent variables, by regarding it as a special case of a more general problem, depending on the larger set of $n+1$ independent variables, and by integrating out the additional variable. 
Observe that we are adopting an informal geometric meaning of the term dimension. In three-dimensional space, three parameters (cartesian coordinates) are required to determine the position of a point. Hadamard simply integrates out one of them to ``descend'' to two-dimensional space, and integrates out a second variable to descend to one-dimensional space. The strength of the method lies in its ingenious simplicity.

The strategy is simple, yet very powerful: low-dimensional EM is viewed as a particular case of the higher dimensional one, in which all relevant quantities are uniform along the additional spatial directions. As we shall see, this procedure, inspired by Hadamard's descent method, will yield two independent EMs in two spatial coordinates, and four independent EMs in one spatial coordinate. Each low-dimensional model will have its own characteristics and physical properties. 
Some of these features have been investigated in relation with the quantization of (lattice) gauge theories in 1 and 2 dimensions \cite{Weinberg,hist1,hist2,hist3,tH74,witten,parisi1975quark}. 
In our analysis we will also recover some models that have been studied in the literature~\cite{Lapidus,MR,Wheeler,Boito,McDonald}.

The interest of EM in lower dimensions is not purely academic. Nowadays quantum technologies enable one to perform quantum simulations of low-dimensional gauge theories~\cite{qsim_book,qsim1,qsim2,qsim3,qsim4,qsim5,qsim6}, enabling one to study phenomena, such as real-time dynamics and the string-breaking mechanism, that were out of the reach of numerical investigation and research until a few decades ago. 

Moreover, recent advances in waveguide quantum electrodynamics make possible the investigation of dimensional reduction, field confinement effects, and field-mediated coupling in one- \cite{KKM,GT11,vL13,F16} and two-dimensional systems \cite{M15,GT15}, enabling comparison among different dimensional features \cite{HSD}. We hope that our analysis can yield insights into these problems.

This article is organized as follows. 
We review electromagnetism in 3+1 dimensions in Sec.\ \ref{3+1}, focusing on its tensor structure. 
We perform Hadamard's first descent, from 3 to 2 spatial dimensions, in Sec.\ \ref{firstdesc}, and observe that 
the Maxwell's equations split up into two independent, uncoupled sets of equations, each one made up of four coupled equations.
We perform the second (and last) descent, from 2 to 1 spatial dimensions, in Sec.\ \ref{seconddesc}: 
the two independent sets of equations obtained from the first descent split up again into two uncoupled sets of equations, yielding four independent electromagnetisms. One of these theories is trivial, and some of them are known in the literature. We conclude in Sec.\ \ref{concl}, with a discussion and a perspective.

\section{Preliminaries: Electromagnetism in 3+1 dimensions}\
\label{3+1}

The starting point of our analysis is the classical formulation of electromagnetism in three spatial dimensions \cite{Jackson,Ingarden,Hehl}.
The tensor structure of the theory is manifest in its formulation in a $(3+1)$-dimensional Minkowski spacetime \cite{Barut}. 
In this work, our main focus is on the behavior of the EM field components as the spatial dimension is lowered; therefore, charges and currents will be considered as non-dynamical sources depending on the space-time coordinates.

\subsection{Fields and tensors}\label{fieldstensors}	

The EM field in the $(3+1)$-dimensional Minkowski space, for which we assume a metric tensor $\eta$ with signature $(\eta^{\mu\nu})=\mathrm{diag}(1,-1,-1,-1)$, can be described in terms of the field tensor $F$,  coupled to a $4$-vector current $j$, and its dual pseudotensor $G$:
	\begin{align}
		\label{eq:emfielddual}
		(F^{\mu\nu}) =
		\left(\begin{BMAT}{cccc}{cccc}
			0 & -E_x & -E_y& -E_z\\
			E_x & 0 & -B_z & B_y\\
			E_y & B_z & 0 &-B_x \\
			E_z & -B_y & B_x& 0
		\end{BMAT}\right) ,
		&& (G^{\mu\nu}) =
		\left(\begin{BMAT}{cccc}{cccc}
			0      & -B_x & -B_y  & -B_z\\
			B_x & 0 	  & E_z   & -E_y\\
			B_y & -E_z & 0 		 & E_x \\
			B_z &  E_y & -E_x & 0
		\end{BMAT}\right) ,
		&&
		(j^{\mu})=\left(\begin{BMAT}{c}{cccc}
			c\rho\\
			J_x\\
			J_y\\
			J_z
		\end{BMAT}\right) .
	\end{align}
In the above equations and throughout the paper, we adopt Heaviside-Lorentz units, in which the components of the electric and magnetic fields have the same dimensions. The tensors $F$ and $G$ are both rank-2 antisymmetric, and duality between them is expressed by the relations $G^{\mu\nu}=\frac{1}{2}\epsilon^{\mu\nu\rho\sigma}F_{\rho \sigma}$,  $F^{\mu\nu}=-\frac{1}{2}\epsilon^{\mu\nu\rho\sigma}G_{\rho \sigma}$, for $\mu,\nu \in \{0,1,2,3\}$, with $\epsilon^{0123}=1$. The components of the electric $\bm E$ and magnetic field $\bm B$, the charge density $\rho$, and the current density $\bm J$, determine the block structure of these quantities, 	\begin{align}\label{eq:SO(3)decomposition}
		(F^{\mu\nu})= 
		\left(\begin{BMAT}{c|c}{c|c}
			& -\bm{E}^\top\\
			\bm{E} & \vphantom{\begin{BMAT}[1pt]{c}{ccc} 0\\0\\0 \end{BMAT}}
			\mathmakebox[\widthof{$\begin{BMAT}[3pt]{ccc}{c}0&0&0\end{BMAT}$}]{-{\bm{\tilde{B}}}}
		\end{BMAT}\right) ,
		&&
		(G^{\mu\nu})=
		\left(\begin{BMAT}{c|c}{c|c}
			& - \bm{B}^\top \\
			\bm{B} & \vphantom{\begin{BMAT}[1pt]{c}{ccc} 0\\0\\0 \end{BMAT}}
			\mathmakebox[\widthof{$\begin{BMAT}[3pt]{ccc}{c}0&0&0\end{BMAT}$}]{\bm{ \tilde{E}}}
		\end{BMAT}\right) ,
		&&(j^{\mu})=\left(\begin{BMAT}{c}{c|c}
			\vphantom{\bm{B}^\top} c\rho\\
			\vphantom{\begin{BMAT}[1pt]{c}{ccc} 0\\0\\0 \end{BMAT}} \bm{J} 
		\end{BMAT}\right) ,
	\end{align}
with $^\top$ denoting transposition, and the tilde representing duality in $\mathbb{R}^3$ ($\bm {\tilde v}$ being the linear map $\bm x\mapsto \bm x\cross\bm v$).
		
Under a proper three-dimensional spatial rotation (in this article we shall not focus on discrete symmetries, such as space and time inversions), both $\bm E$ and $\bm B$ transform like 3-vectors, no mixing between the electric and magnetic components occurs, and the block structure~\eqref{eq:SO(3)decomposition} of $F$ and $G$ is preserved, as expected from a general property of rank-2 antisymmetric Minkowski tensors, whose $3\oplus 3$ 
decomposition is invariant with respect to proper rotations $\SO$. Likewise, proper rotations leave the charge density $\rho$ unchanged, and transform the current density $\bm J$ like a 3-vector, preventing mixing between $\rho$ and the components of $\bm J$ and preserving the block structure~\eqref{eq:SO(3)decomposition} of $j$, that thus behaves like a vector, characterized by an invariant $1\oplus 3$ decomposition with respect to $\SO$.

The dynamics in an inertial reference frame is determined by the Maxwell's equations
	\begin{align}
		\partial_\mu G^{\mu\nu}&=0,							\label{eq:maxwelltensorhomogeneous}\\
		\partial_\mu F^{\mu\nu}&=\frac{1}{c}j^\nu, \label{eq:maxwelltensorinhomogeneous}
	\end{align}
with $(\partial_{\mu})=(c^{-1}\partial_t,\partial_x,\partial_y,\partial_z)$, that read, in terms of the electric and magnetic fields,
\begin{flalign}
		&\text{\sc (magnetic Gauss)}&\Div \bm B&=0 ,					&&&\\
		&\text{\sc (Faraday)}& \frac{1}{c}\partial_t\bm B+ \Curl \bm E
		&=\bm 0,													\label{eq:faraday}&&&\\
		&\text{\sc (electric Gauss)}&\Div \bm E&= \rho ,					&&&\\
		&\text{\sc (Amp\`{e}re-Maxwell)}& -\frac{1}{c}\partial_t\bm E+\Curl \bm B
		&=\frac{1}{c}\bm J.										\label{eq:ampere-maxwell}&&&
	\end{flalign}
It will be convenient in the following to have the explicit expression of the above Maxwell's equations in components:
\begin{align}
    \partial_x B_x+\partial_y B_y + \partial_z B_z &= 0, \label{eq:magneticgauss}\\
    \frac{1}{c}\partial_tB_x +\partial_y E_z - \partial_z E_y&=0,	\label{eq:faradayx} \\
    \frac{1}{c}\partial_tB_y +\partial_z E_x - \partial_x E_z&=0,	\label{eq:faradayy} \\
	\frac{1}{c}\partial_tB_z +\partial_x E_y - \partial_y E_x&=0,	\label{eq:faradayz} \\
	\partial_xE_x+\partial_y E_y + \partial_z E_z &= \rho, \label{eq:electricgauss}\\
	-\frac{1}{c}\partial_tE_x + \partial_y B_z - \partial_z B_y &=\frac{1}{c}J_x,  \label{eq:ampere-maxwellx} \\
	-\frac{1}{c}\partial_tE_y + \partial_z B_x - \partial_x B_z &=\frac{1}{c}J_y,\label{eq:ampere-maxwelly} \\
	-\frac{1}{c}\partial_tE_z + \partial_x B_y - \partial_y B_x &=\frac{1}{c}J_z .\label{eq:ampere-maxwellz}
\end{align} 

Starting from Maxwell's equations, one can derive a wave equation for each of the EM field components
	\begin{equation}\label{eq:wavepropagation}
		\Box F^{\mu\nu}=\frac{1}{c}\qty(\partial^\mu j^\nu-\partial^\nu j^\mu),
	\end{equation}
with 	$\Box = \partial^\mu\partial_\mu = c^{-2}\partial_t^2 - \partial_x^2 - \partial_y^2 - \partial_z^2 = c^{-2}\partial_t^2-\Delta$, where the role of the sources is played by the antisymmetrized derivatives of the 4-current components.

\subsection{Potentials and Lagrangian}
\label{potl}
The homogeneous Maxwell's equations~(\ref{eq:maxwelltensorhomogeneous}), that can be equivalently expressed as
\begin{equation}
\epsilon^{\mu\nu\rho\sigma} \partial_{\nu} F_{\rho \sigma} = 0,
\end{equation}
provide source-independent constraints on the components of the EM field. These equations are satisfied by expressing the field tensor in terms of a $4$-potential vector $A=(A^{\mu})=(\Phi,A_x,A_y,A_z)$, such that 
	\begin{equation}\label{eq:fieldpotentialrelations}
		F^{\mu\nu}=\partial^\mu A^\nu-\partial^\nu A^\mu,
	\end{equation}
for $\mu,\nu\in\{0,1,2,3\}$. In the non-covariant formalism, the relation between fields and potentials reads
\begin{align}
	\bm B&=\Curl \bm A,\label{eq:classicalpotentiali}\\
	\bm E&=-\grad\varPhi-\frac{1}{c}\partial_t\bm A .
\end{align}
In this way, the number of degrees of freedom of the EM field is reduced to four. A further reduction derives from the observation that the definition of fields in terms of potentials is invariant under gauge transformations
\begin{equation}
A_{\mu} \to {A'}_{\mu} = A_{\mu} + \partial_{\mu} f,
\end{equation}
with $f=f(t,x,y,z)$ being a scalar function. If the gauge choice is made in such a way that the Lorentz condition $\partial_\nu A^\nu=0$ holds, then the potential obeys the wave equation 
\begin{equation}
\label{eq:awave}
\dal A^\mu=\frac{1}{c} j^\mu ,
\end{equation}
from which the wave equation~(\ref{eq:wavepropagation}) for the field components can be immediately derived.

While the homogeneous equations are a direct consequence of expressing $F^{\mu\nu}$ as an antisymmetrized derivative of the potential, the inhomogeneous equations~\eqref{eq:maxwelltensorinhomogeneous} can be derived from the Lagrangian density
\begin{align}
\label{eq:L3D}	
		\mathcal{L}_{3+1}(A) & = -\frac{1}{4}(\partial^\mu A^\nu-\partial^\nu A^\mu)(\partial_\mu A_\nu-\partial_\nu A_\mu)-\frac{1}{c}j^\mu A_\mu \nonumber \\
		& =\frac{1}{2}\qty(\bm E^2-\bm B^2)- \rho\,\Phi+\frac{1}{c} \bm J\cdot\bm A ,
	\end{align}
by considering the set of Euler-Lagrange equations associated to each component $A^{\mu}$ of the potential.

\section{First descent: from 3 to 2 spatial dimensions}
\label{firstdesc}			
	\subsection{Fields}
	\label{1descent}	

\begin{figure}
		\centering
		\begin{subfigure}[b]{0.48\textwidth}
			\includegraphics[width=\linewidth]{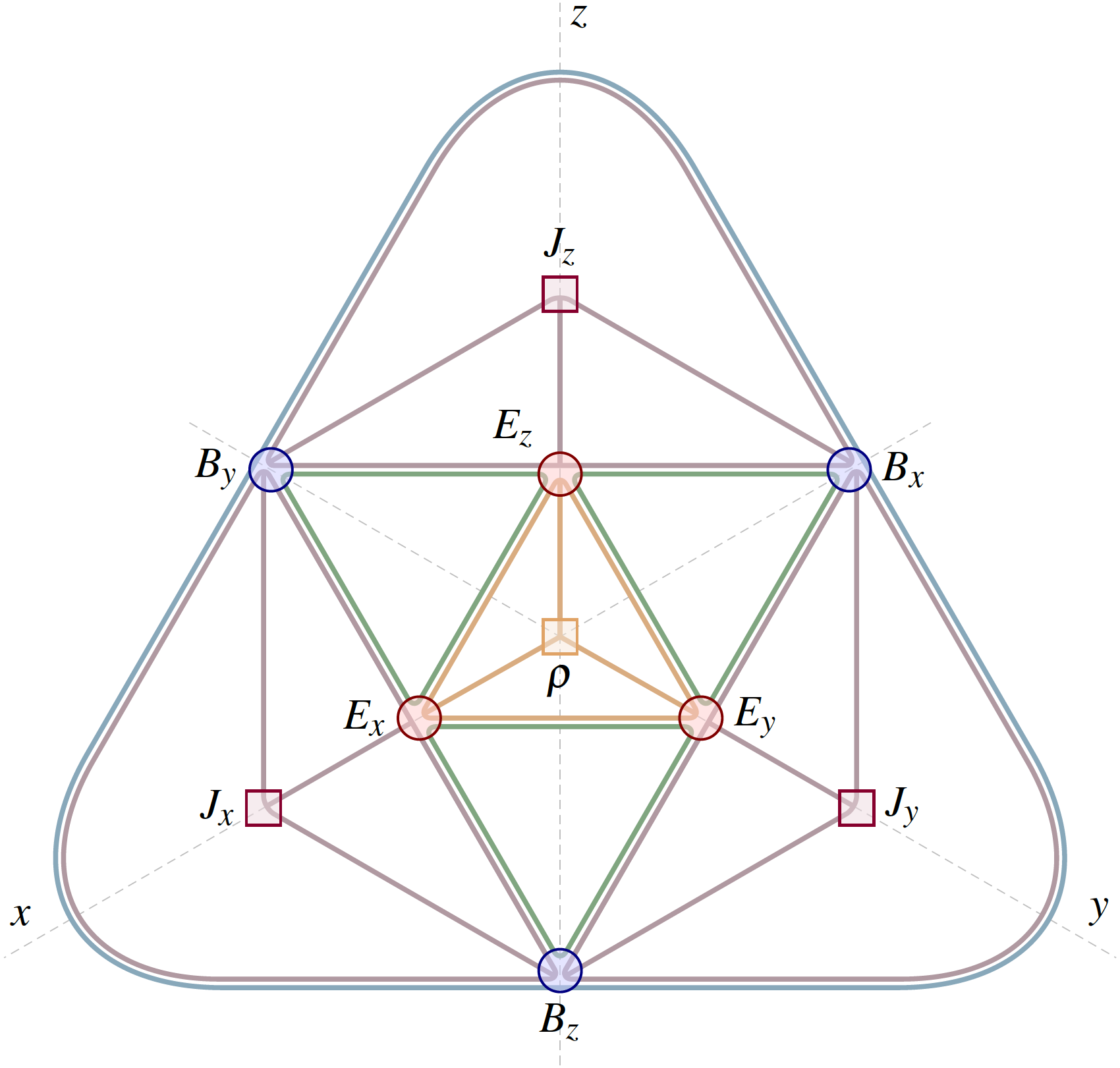}
			\caption{} \label{fig:maxwell}
		\end{subfigure}\hfill
		\begin{subfigure}[b]{0.48\textwidth} 
			\centering
			\includegraphics[width=\linewidth]{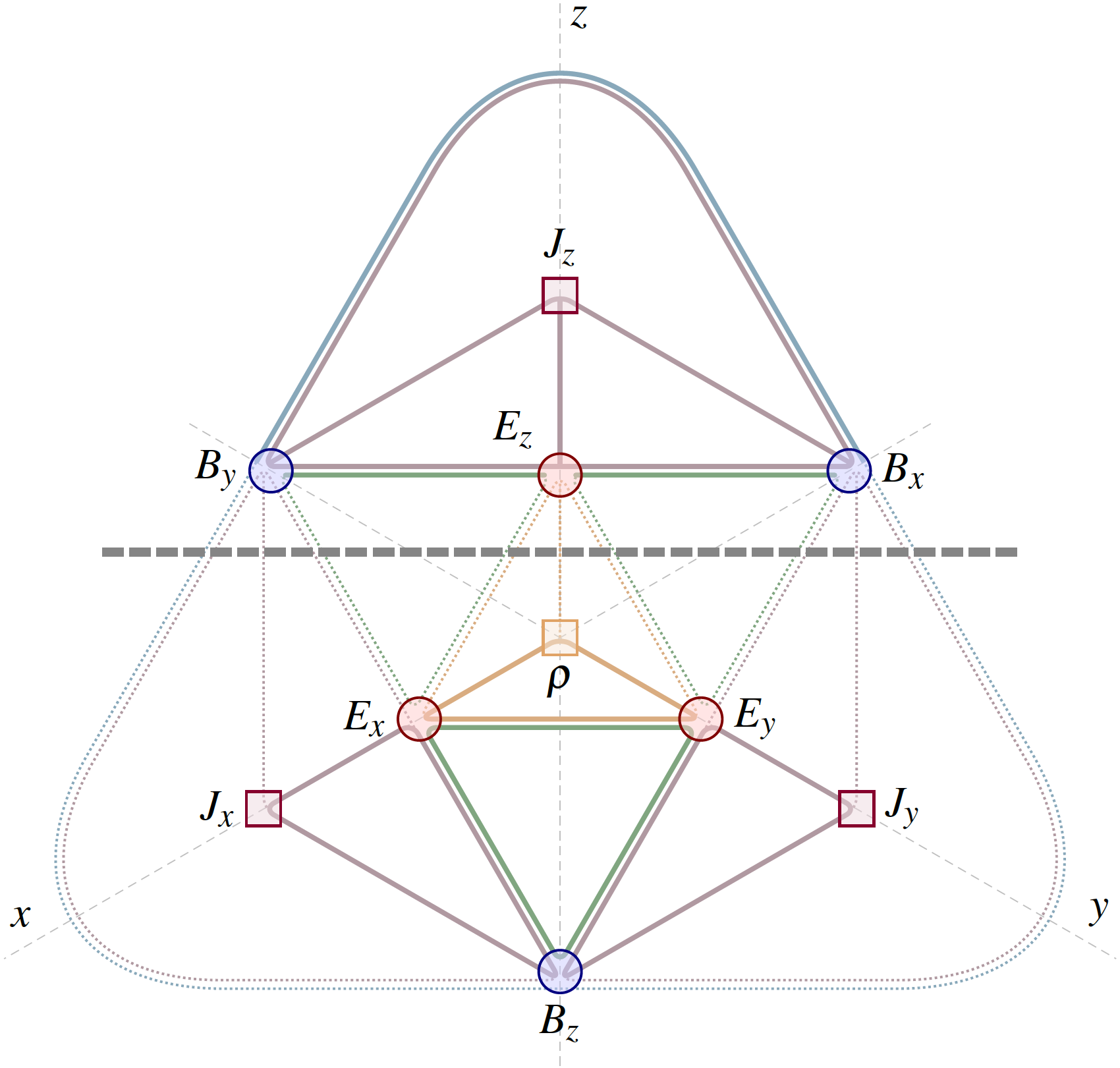}
			\caption{} \label{fig:discesa}
		\end{subfigure}	
		\caption{\raggedright Graph representation of (a) the structure of the Maxwell's equations in $3+1$ dimensions, and (b) after the first dimensional reduction. Vertices are either fields ($\bm E$ (\protect\electric), $\bm B$~(\protect\magnetic)) or sources ($\rho$~(\protect\charge), $\bm J$~(\protect\current)), and edges show the couplings, two vertices being linked if and only if they appear in the same differential equation ({\sc electric Gauss}~(\protect\gaussel), {\sc magnetic Gauss}~(\protect\gaussmag), {\sc Faraday}~(\protect\faraday), {\sc Amp\`{e}re-Maxwell}~(\protect\ampere)). The assumption of $z$-independence is represented by the dashed horizontal line, that isolates two sectors governed by independent field equations. Maxwell system in $3+1$ dimensions is a connected graph, and each equation is a connected subgraph. By performing the descent, the graph splits up into two connected components: the homogeneous equations~(\ref{eq:BxByEzi})-(\ref{eq:ByEz}) are reduced to a single edge, the inhomogeneous equations~(\ref{eq:ExEyBzi})-(\ref{eq:ExBz}) to a triangular subgraph, whereas Eqs.~(\ref{eq:ExEyBzi}) and~(\ref{eq:BxByEzf}) (which are also the only equations involving all the variables of each set) are unchanged.}
		\label{fig:maxwellDiscesa}
	\end{figure}

We now apply Hadamard's strategy and perform the first descent along the $z$ direction, specializing the Maxwell differential problem into a $z$-independent one (any other direction would lead to identical results).
We start from Maxwell's equations in 3+1 dimensions and require that both sources and solutions are $z$-independen, so that all terms containing derivatives with respect to $z$ vanish. It is straightfoward to see that Maxwell's equations (\ref{eq:magneticgauss})-(\ref{eq:ampere-maxwell})
split up into two independent (uncoupled) sets, each one made up of 4 equations. We give all details in the following.
See Fig.~\ref{fig:maxwellDiscesa}.

\subsubsection{$(E_x , E_y, B_z)$ sector}		
The first subsystem is obtained from Eqs.~\eqref{eq:faradayz}, \eqref{eq:electricgauss}, \eqref{eq:ampere-maxwellx}, and \eqref{eq:ampere-maxwelly}.
It determines the dynamics and the constraints involving the field components $E_x $, $E_y$, and $ B_z$,
\begin{align}
	\frac{1}{c}\partial_tB_z +\partial_x E_y - \partial_y E_x&=0,				\label{eq:ExEyBzi}\\
	\partial_xE_x+\partial_y E_y&= \rho,\\
	-\frac{1}{c}\partial_tE_x + \partial_yB_z&=\frac{1}{c}J_x,  \label{eq:ExBz} \\
	-\frac{1}{c}\partial_tE_y -\partial_x B_z&=\frac{1}{c}J_y,\label{eq:ExEyBzf}
\end{align} 
and contains three inhomogeneous equations, with source terms $\rho$, $J_x$, and $J_y$. Observe that, since $\partial_z J_z=0$, the continuity equation, $\partial_t \rho+\partial_xJ_x+\partial_yJ_y=0$, pertains only to this subsystem. This is the usual reduction of EM in 2+1 dimensions, that preserves the minimal features of the 3+1 structure. We propose a pictorial representation in Fig.\ \ref{fig:3Dto2Da}.

\subsubsection{$(B_x , B_y, E_z)$ sector}
\begin{figure}[h!t]
	\centering
	\begin{subfigure}[b]{.495\textwidth}
		\includegraphics[width=\textwidth]{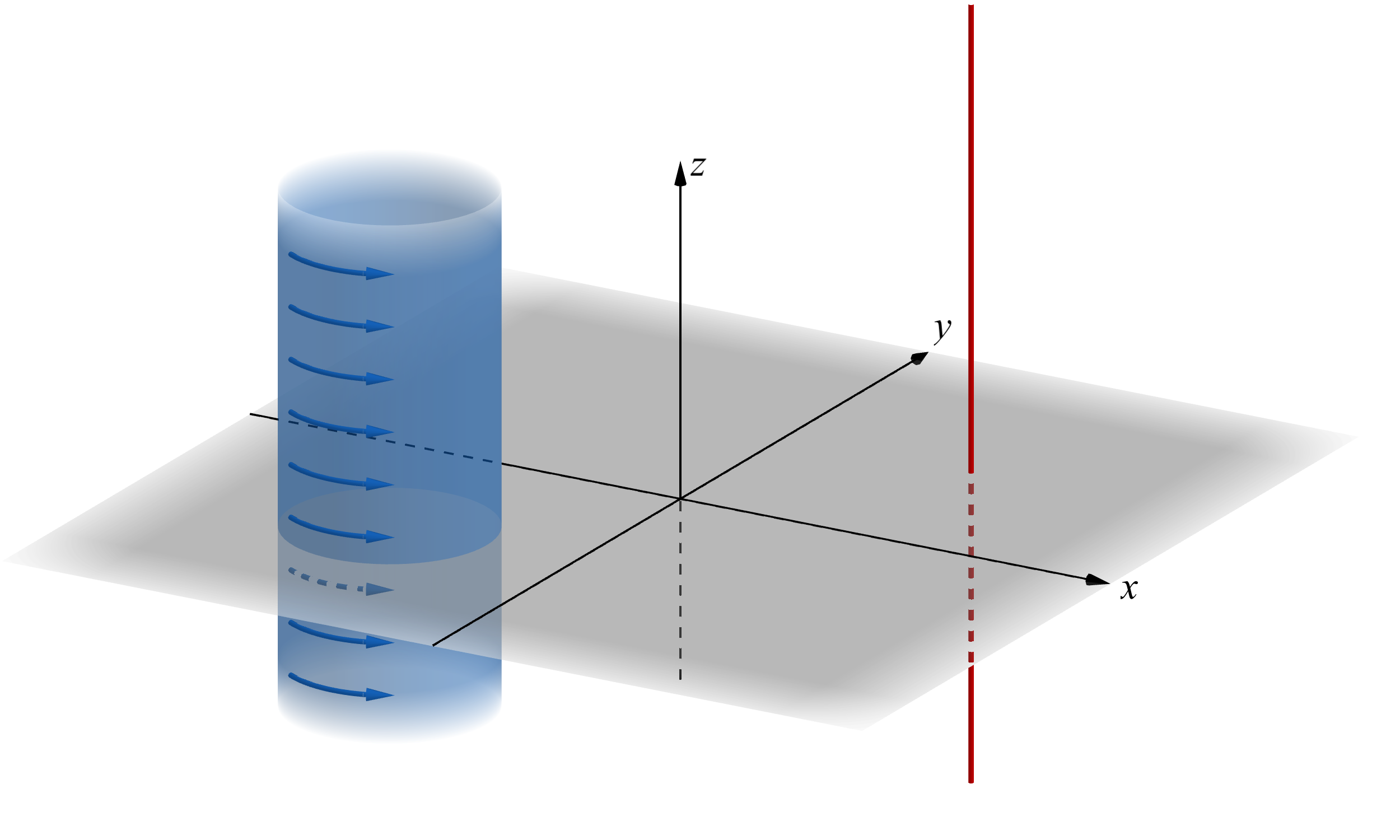}	
		\caption{} \label{fig:3Dto2D_A}
		\includegraphics[width=\textwidth]{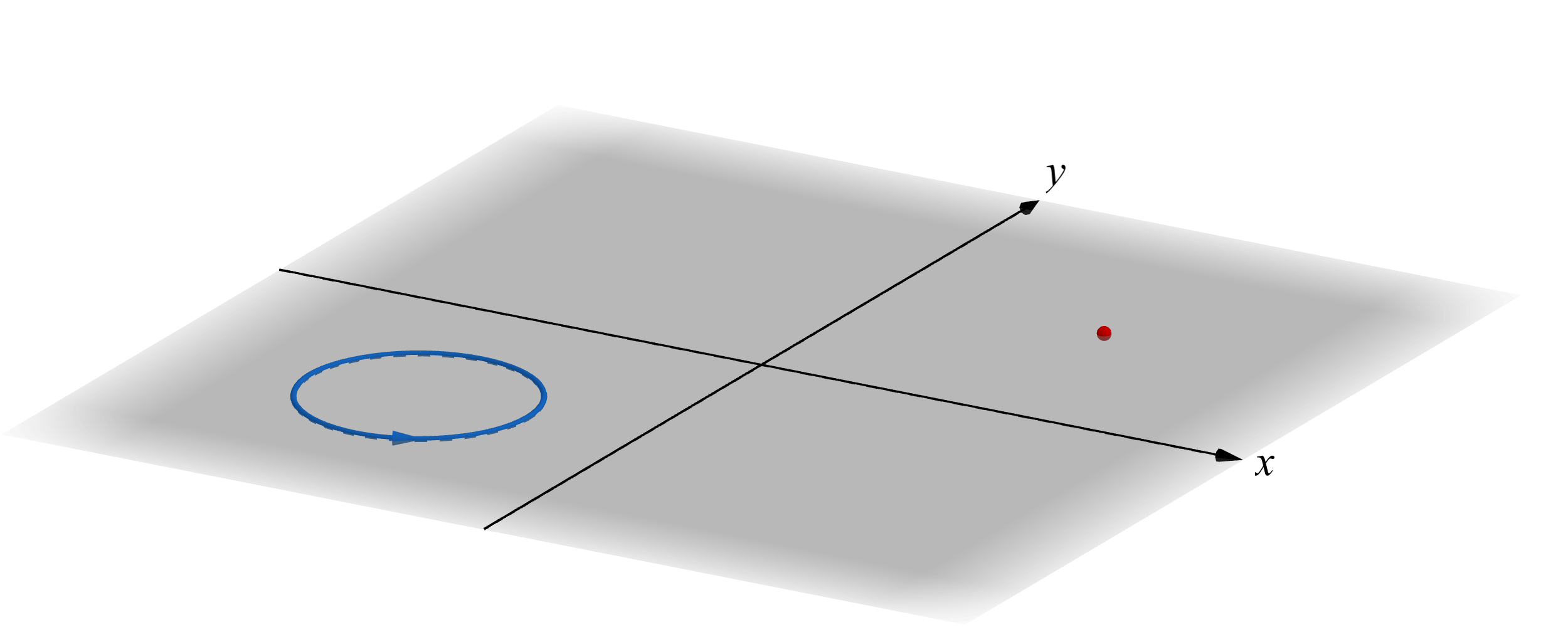}	
		\caption{} \label{fig:3Dto2D_B}
	\end{subfigure}	\hfill
	\begin{subfigure}[b]{.495\textwidth}
		\includegraphics[width=\textwidth]{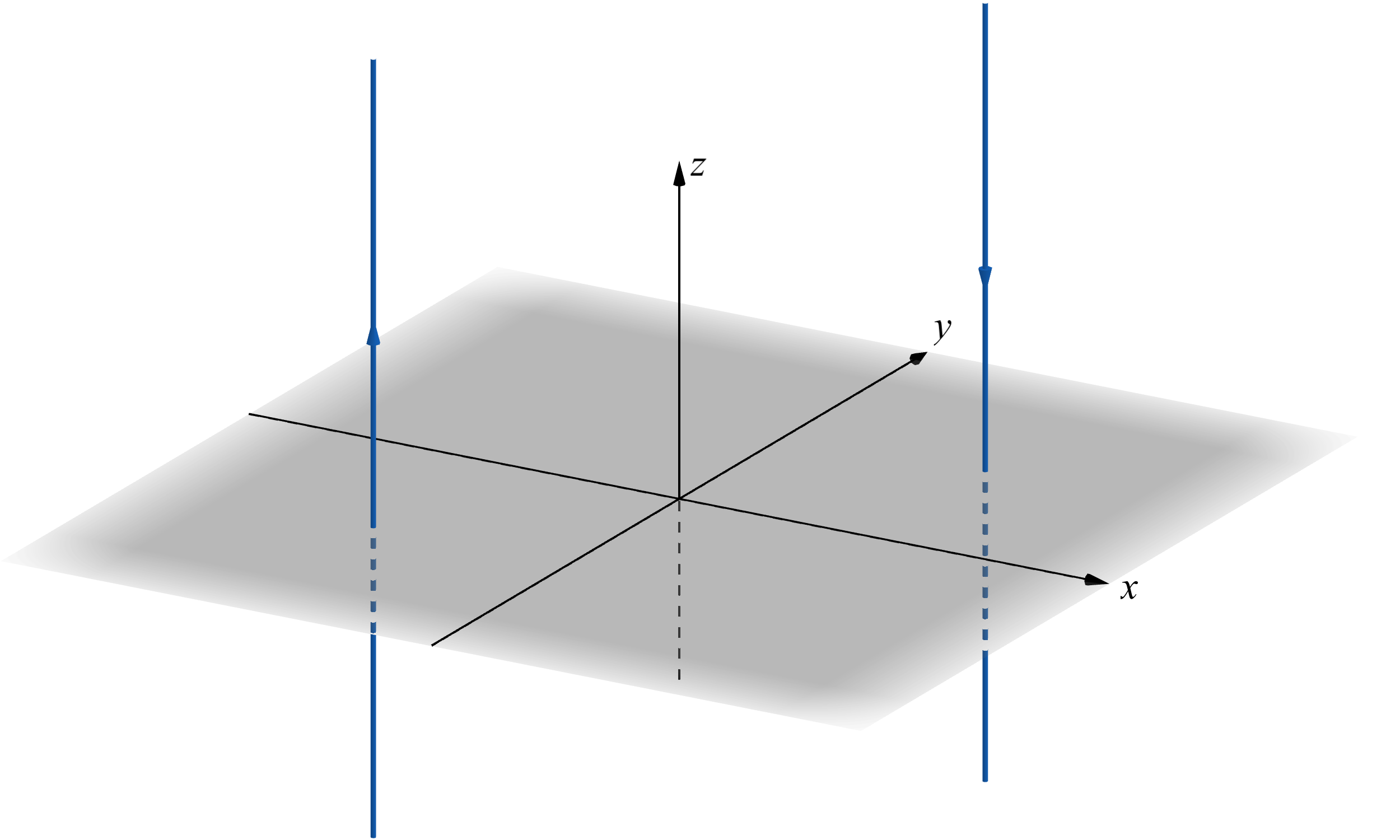}	
		\caption{} \label{fig:3Dto2D_C}
		\includegraphics[width=\textwidth]{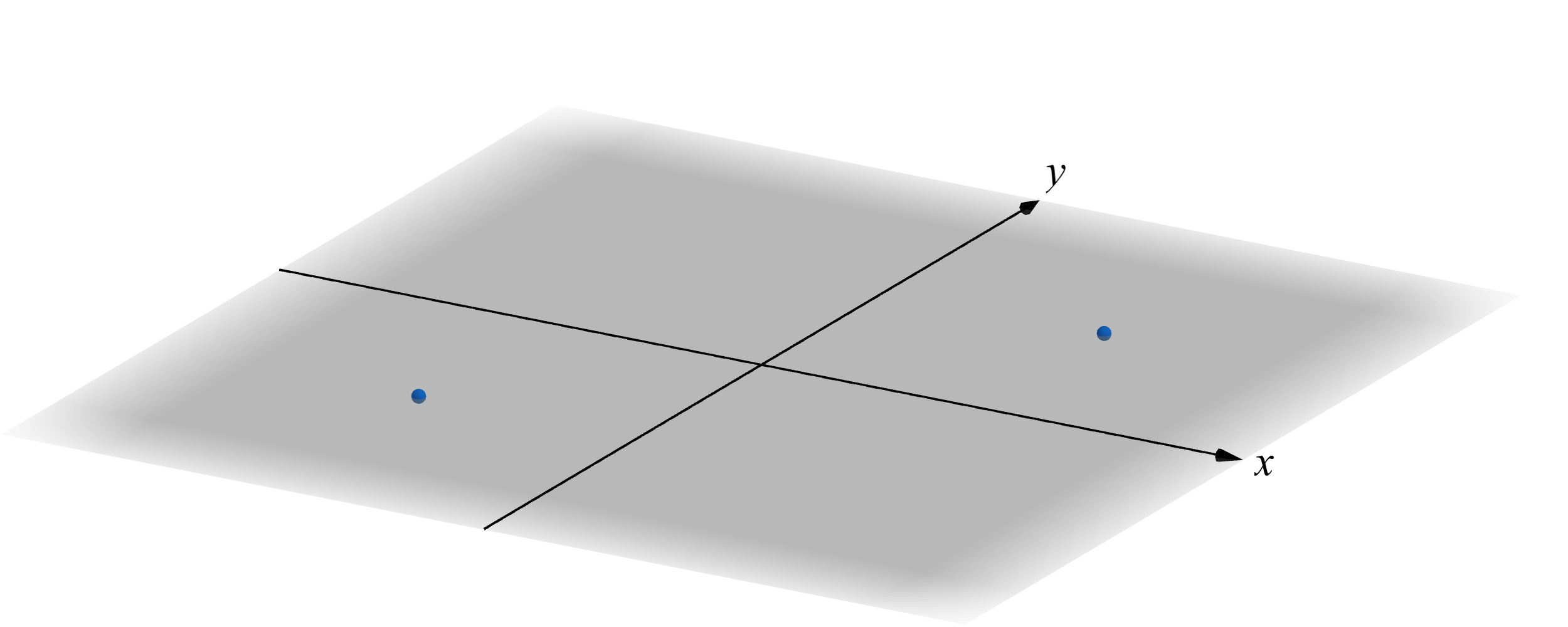}	
		\caption{} \label{fig:3Dto2D_D}
	\end{subfigure}	
	\caption{\raggedright Top panels: $z$-independent sources in a three-dimensional space. Lower panels: their representation in $(2+1)$ electromagnetism. $(E_x,E_y,B_z)$ sector: (a) an indefinite (red) charged wire and a (blue) solenoid, with current vector orthogonal to $z$ (b) are represented in two dimensions by a (red) point charge and a (blue) circular current loop in the $(x,y)$ plane. $(B_x,B_y,E_z)$ sector: (c) two indefinite (blue) current wires parallel to $z$ (d) are represented in two dimensions by two blue ``point charges'' of opposite sign.}
	\label{fig:3Dto2Da}
\end{figure}

The second subsystem derives from Eqs.~\eqref{eq:magneticgauss}, \eqref{eq:faradayx}, \eqref{eq:faradayy}, and \eqref{eq:ampere-maxwellz}, and involves the complementary set of field components. Maxwell's equations in this sector reduce to
\begin{align}
	\partial_xB_x+\partial_y B_y&=0,\label{eq:BxByEzi}  \\
	\frac{1}{c}\partial_tB_x + \partial_y E_z&=0,\\
	\frac{1}{c}\partial_tB_y -\partial_x E_z&=0, \label{eq:ByEz} \\
	-\frac{1}{c}\partial_tE_z +\partial_x B_y - \partial_y B_x&=\tfrac{1}{c}J_z,
	\label{eq:BxByEzf}
\end{align} 
where a single inhomogeneous equation is present, with source term $J_z$. See again Fig.~\ref{fig:3Dto2Da}.

\subsection{Tensor notation}\label{1descenttensor}	
As a result of the first descent, the EM field tensor, its dual and  the $4$-current are partitioned as follows:
	\begin{align}
		\label{eq:tensordecomposition}
		(F^{\mu\nu})=\left(
		\begin{BMAT}{ccc1c}{ccc1c}
			0 & \textcolor{red!75!black}{-E_x} & \textcolor{red!75!black}{-E_y} & \textcolor{blue!75!black}{-E_z}\\
			\textcolor{red!75!black}{E_x} & 0 & \textcolor{red!75!black}{-B_z} & \textcolor{blue!75!black}{B_y}\\
			\textcolor{red!75!black}{E_y} & \textcolor{red!75!black}{B_z} & 0 &\textcolor{blue!75!black}{-B_x}\\
			\textcolor{blue!75!black}{E_z} & \textcolor{blue!75!black}{-B_y}& \textcolor{blue!75!black}{B_x}  & 0
		\end{BMAT}\right),
		&&
		(G^{\mu\nu})=\left(
		\begin{BMAT}{ccc1c}{ccc1c}
			0& \textcolor{blue!75!black}{-B_x} & \textcolor{blue!75!black}{-B_y} & \textcolor{red!75!black}{-B_z}\\
			\textcolor{blue!75!black}{B_x} & 0 & \textcolor{blue!75!black}{E_z} & \textcolor{red!75!black}{-E_y}\\
			\textcolor{blue!75!black}{B_y} & \textcolor{blue!75!black}{-E_z} & 0 & \textcolor{red!75!black}{E_x}\\ 
			\textcolor{red!75!black}{B_z} & \textcolor{red!75!black}{E_y} &\textcolor{red!75!black}{E_x} & 0
		\end{BMAT}\right),
		&&
		(j^{\mu})=\left(\begin{BMAT}{c}{ccc1c}
			\textcolor{red!75!black}{c\rho}\\
			\textcolor{red!75!black}{J_x}\\
			\textcolor{red!75!black}{J_y}\\
			\textcolor{blue!75!black}{J_z}
		\end{BMAT}\right),
	\end{align}	
where the components related to the first subsystem are in blue, and those related to the second one in red. 
By switching to tensor notation, we see that the manifest duality between the blocks of equal color in $F$ and $G$ is a consequence of the interplay between $(3+1)$- and $(2+1)$-duality for a rank-2 antisymmetric $(3+1)$-tensor:
	\begin{align}\label{eq:duality}
		G^{ab}
		&=\frac{1}{2}\qty(\epsilon^{{abc}3}F_{{{c}}3}+\epsilon^{{ab}3 c}F_{3{c}})=
		\epsilon^{{abc}}F_{{c}3} ,\\
		\label{eq: 3+1 vs 2+1 duality 2}
		G^{{a} 3}
		&=\frac{1}{2}\epsilon^{{a} 3 {bc}}F_{{bc}}
		=\frac{1}{2}\epsilon^{{abc}}F_{{bc}},
	\end{align}
Henceforth, the latin indices $a,b,c$ are assumed to vary in $\{0,1,2\}$, regardless of whether they are free or contracted. In tensor form, the system of equations~\eqref{eq:ExEyBzi}--\eqref{eq:ExEyBzf} 
of the $(E_x , E_y, B_z)$ sector reads
\begin{align}
	\epsilon^{{abc}}\partial_{{a}}  F_{{bc}}&=0,							\label{eq:homFab}\\
	\partial_{{a}} F^{{ab}}&=\frac{1}{c} j^{{b}},										\label{eq:inhomFab}
\end{align}
with ${a}\in\{0,1,2\}$. $(F^{{ab}})$ is an antisymmetric tensor of order $3$, while its dual $(G^{{a}3})$ behaves as a vector in a $(2+1)$-dimensional space, like the source $(j^{{a}})$. 

The tensor form of the Maxwell's equations~\eqref{eq:BxByEzi}--\eqref{eq:BxByEzf} for the $(B_x,B_y,E_x)$ sector reads
\begin{align}
	\epsilon^{{abc}}\partial_{{b}}  F_{{c}3}&=0,      						\label{eq:homFa3}\\
	\partial_{{a}} F^{{a}3}&=\frac{1}{c} j^3,														\label{eq:inhomFa3}
\end{align}
where $(F^{a3})$ is now a $3$-component vector, its dual $(G^{{ab}})$ an antisymmetric tensor of order $3$, and the source $j^3$ is a scalar. The form of Eqs.~\eqref{eq:homFab}--\eqref{eq:inhomFa3} is similar to that of the relativistic Maxwell's equations~\eqref{eq:maxwelltensorhomogeneous}--\eqref{eq:maxwelltensorinhomogeneous}. However, notice that, in each subsystem, the field tensor and the source are different kinds of  objects, with a closer analogy to the $(3+1)$ case valid only in the $(E_x,E_y,B_z)$ sector.

In both sectors, Maxwell's equations determine that field components propagate as waves in two dimensions: indeed, in a $z$-independent framework, the d'A\-lem\-bert operator is effectively replaced by $\partial_{a}\partial^{a} = c^{-2}\partial_t^2 - \partial_x^2-\partial_y^2$, and we obtain from Eq.~\eqref{eq:wavepropagation}
\begin{align}
\partial_{a}\partial^{a} F^{{bc}} & =\frac{1}{c}(\partial^{{b}} j^{{c}}-\partial^{{c}} j^{{b}}), \label{eq:32}\\
\partial_{{a}}\partial^{{a}}  F^{{{b}} 3} & =\frac{1}{c}\partial^{{b}} j^3. \label{eq:33}
\end{align}
Notice, also in this case, the different form of the equation~\eqref{eq:33} in the $(B_x,B_y,E_z)$ sector, highlighting the effective scalar nature of the source.

\subsection{Remark: stationary electromagnetism as a special case of descent}\label{intinter}	

\begin{figure}
		\centering
		\begin{subfigure}[b]{0.45\textwidth}
			\centering
			\includegraphics[width=\linewidth]{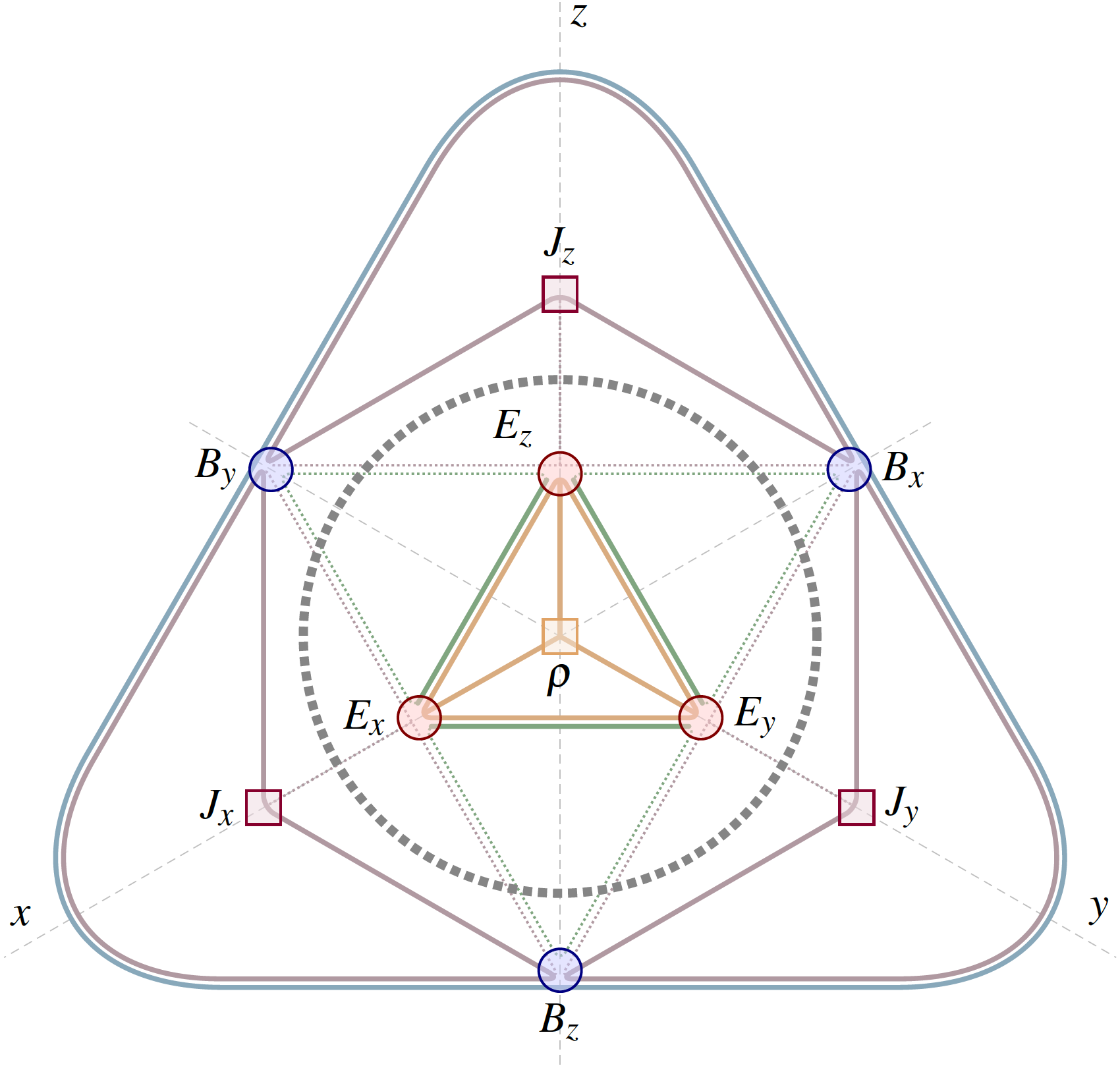}
			\caption{} \label{fig:stationario}
		\end{subfigure}\hfill
		\begin{subfigure}[b]{0.2211\textwidth}
			\centering
			\includegraphics[width=\linewidth]{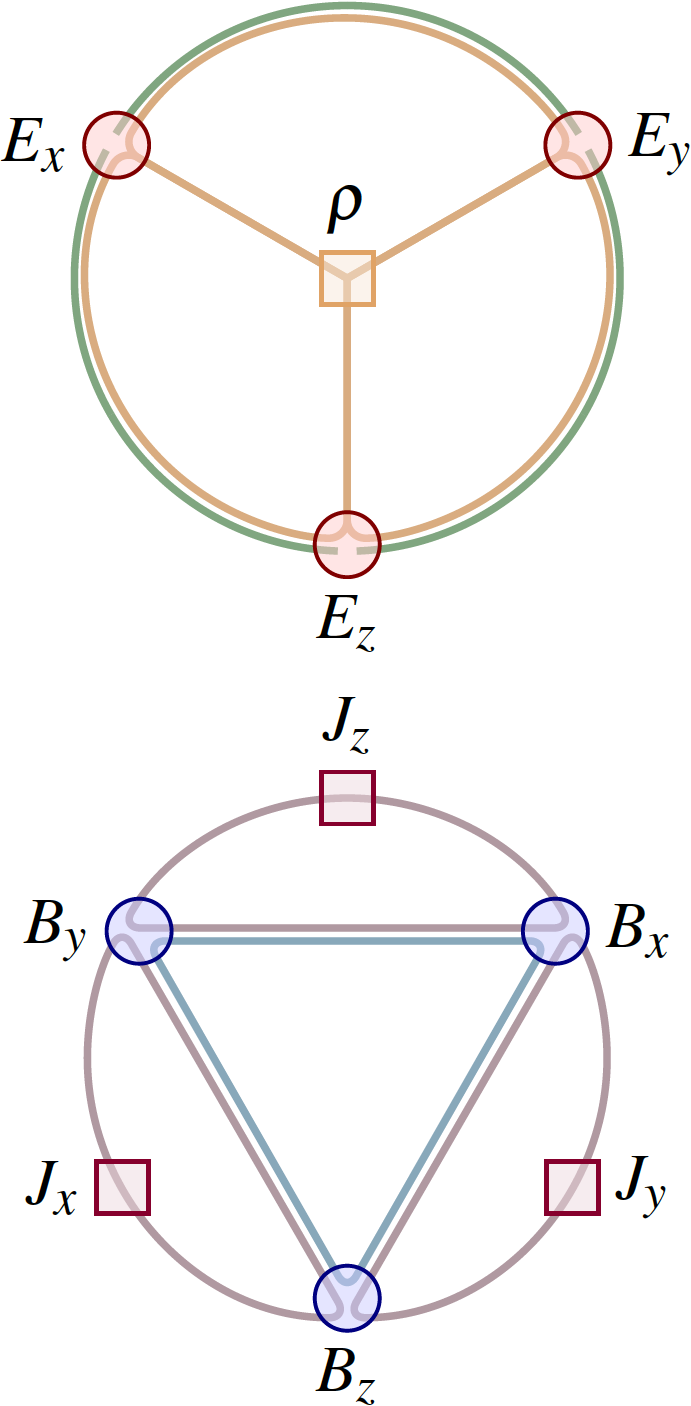}
			\caption{} \label{fig:stationarioschema}
		\end{subfigure}
		\hfill
		\begin{subfigure}[b]{0.2211\textwidth}
			\centering
			\includegraphics[width=\linewidth]{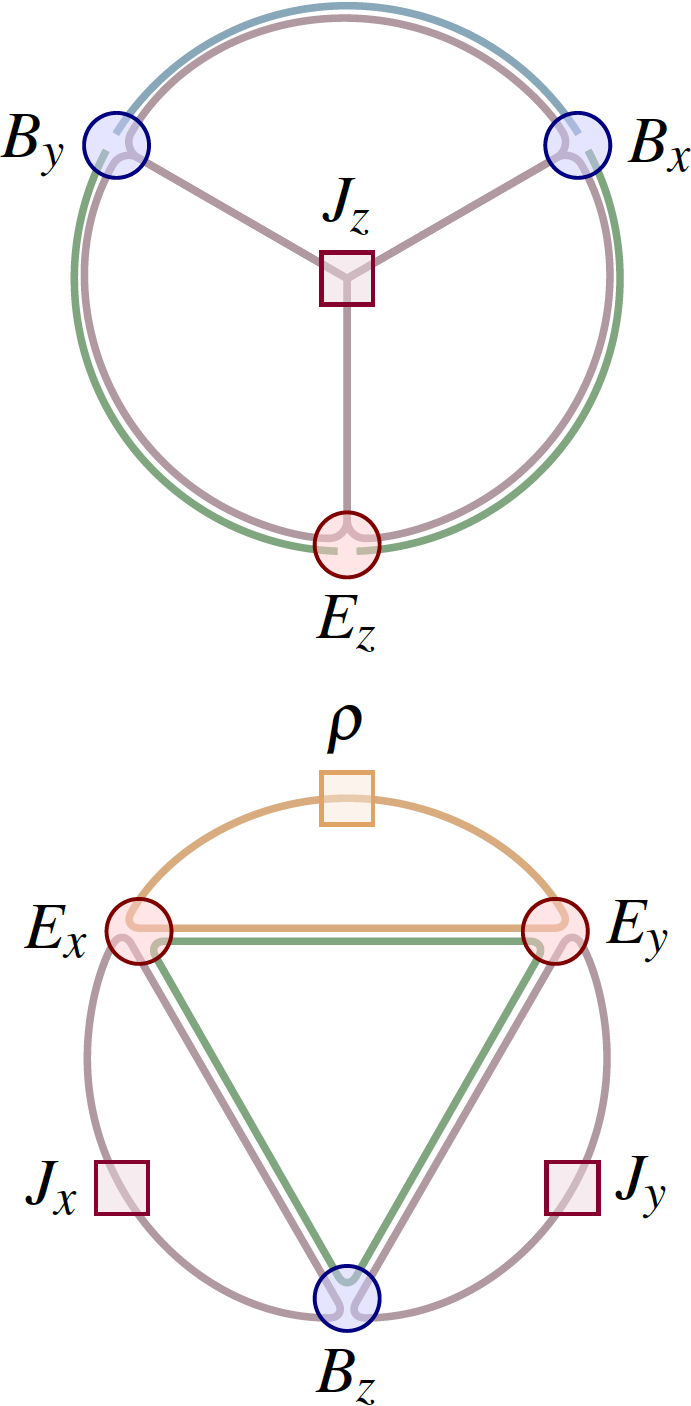}
			\caption{} \label{fig:discesaschema}
		\end{subfigure}
		\caption{\raggedright Graph representation of the stationary Maxwell's equations [panels (a)-(b)], compared with the result of the reduction along the third space direction [panel (c)]. 		
		Time-independence is represented by the dashed circle in panel (a), isolating $\bm B$~(\protect\magnetic) and $\bm J$~(\protect\current) from $\bm E$~(\protect\electric) and $\rho$~(\protect\charge). While {\sc electric Gauss}~(\protect\gaussel) and {\sc magnetic Gauss}~(\protect\gaussmag) are preserved, each component of {\sc Faraday}~(\protect\faraday) is reduced to a single edge, and each components of {\sc Amp\`{e}re-Maxwell}~(\protect\ampere) to a triangular subgraph. The effect is similar to what happens in the descent along $z$.}	\label{fig:stationariovsdiscesa}
	\end{figure}

In the previous subsections, we derived the equations for the electromagnetic field in the case of invariance along a given spatial coordinate. It is worth remarking that even the familiar magnetostatics, 
\begin{align}
				\Div \bm B&=0 ,									\\
	\Curl \bm B
		&=\frac{1}{c}\bm J,
	\end{align}
and electrostatics, 
\begin{align}
				\Curl \bm E
		&=\bm 0,	\\
		\Div \bm E&= \rho ,											
		\end{align}
can be viewed as special cases of Maxwell's equations obtained by dimensional reduction, performed in this case along the time axis $t$. Indeed, the condition of time independence of the fields generates two sectors, the magnetostatic one, in which the components $(B_x,B_y,B_z)$ appear in the equations
\begin{align}
	\epsilon^{k\ell m}\partial_{k}  F_{\ell m}&=0, \label{eq:34}\\
	\partial_{k} F^{k\ell}&=\frac{1}{c} j^{\ell}, \label{eq:35}	
\end{align}
with $k,\ell,m\in\{1,2,3\}$, and the electrostatic one, in which the components $(E_x,E_y,E_z)$ satisfy
\begin{align}
	\epsilon^{k\ell m}\partial_{k}  F_{\ell 0}&=0, \label{eq:36} \\
	\partial_{k} F^{k 0}&= \frac{1}{c} j^0 . \label{eq:37}
\end{align}
The correspondence between the magnetostatics~\eqref{eq:34}--\eqref{eq:35} and the $(E_x , E_y, B_z)$ EM~\eqref{eq:homFab}--\eqref{eq:inhomFab} on one hand, and between the electrostatics~\eqref{eq:36}--\eqref{eq:37} and the $(B_x,B_y,E_x)$ EM~\eqref{eq:homFa3}--\eqref{eq:inhomFa3} is fairly evident. The derivation of static EM and the parallelism between the two cases of space and time descent is graphically represented in Fig.~\ref{fig:stationariovsdiscesa}. 
However, due to the signature of the Minkowski metric tensor, the dimensional reduction along the time axis reduces the  d'A\-lem\-bert operator to the Laplacian $\partial_{k}\partial^{k} = - \Delta$, and instead of wave propagation~\eqref{eq:32}--\eqref{eq:33} we have the Poisson equations
\begin{align}
-\Delta F^{{k\ell}} & =\frac{1}{c}(\partial^{k} j^{\ell}-\partial^{\ell} j^{k}),\\
-\Delta  F^{{{k}} 0} & =\frac{1}{c}\partial^{{k}} j^0 .
\end{align}

\subsection{Potentials and Lagrangian}\label{potl2}

The two sectors emerging from the descent along $z$ are related to two distinct parts of the $(3+1)$ EM Lagrangian density
\begin{equation}
    \mathcal{L}_{3+1} = \left(-\frac{1}{4}F_{{ab}}F^{{ab}}-\frac{1}{c}j^{a} A_{a}\right) + \left(-\frac{1}{2}F_{{a}3}F^{{a}3}-\frac{1}{c}j^3 A_3\right) =: \mathcal{L}_{EEB} + \mathcal{L}_{BBE}.
\end{equation}
The inhomogeneous equations~\eqref{eq:inhomFab} and~\eqref{eq:inhomFa3} can be obtained from the Euler-Lagrange equations generated by the single terms $\mathcal{L}_{EEB}$ and $\mathcal{L}_{BBE}$, respectively, specialized to the case of $z$-independent fields and sources. Notice, however, that in general the two parts of the Lagrangian are not decoupled, since the components $A_{{a}}$ appear not only in $\mathcal{L}_{EEB}$, but also in $\mathcal{L}_{BBE}$, through their derivatives $\partial_3 A_{{a}}$ along the descent coordinate. However, decoupling can occur even at the level of the Lagrangian, by choosing a $z$-independent potential, as for the fields and the sources. The gauge fixing $\partial_3 A_{{a}}=0$, analogous to the request of time-independent potentials in stationary EM, and evidently allowed by both the homogeneous and the inhomogeneous equations~\eqref{eq:homFa3}--\eqref{eq:inhomFa3}, decouples the total Lagrangian density into the sum of the terms
\begin{align}
    \mathcal{L}_{EEB}(A^0,A^1,A^2) & = -\frac{1}{4} (\partial_{{a}}A_{{b}} - \partial_{{b}}A_{{a}})(\partial^{{a}}A^{{b}} - \partial^{{b}}A^{{a}}) -\frac{1}{c}j^{a} A_{a} \nonumber \\
    & = \frac{1}{2} (E_x^2+E_y^2-B_z^2) - \rho\, \Phi + \frac{1}{c} (J_x A_x + J_y A_y)
\end{align}
and
\begin{equation}
    \mathcal{L}_{BBE}(A^3) = -\frac{1}{2} (\partial_{{a}}A_3) (\partial^{{a}}A^3) -\frac{1}{c}j^3 A_3 = \frac{1}{2} (E_z^2-B_x^2-B_y^2) + \frac{1}{c} J_z A_z ,
\end{equation}
leaving in the first sector the gauge freedom with respect to transformations $A_{{a}}\to A_{{a}} + \partial_{{a}}f$, with $f=f(t,x,y)$. The Euler-Lagrange equation in a $(2+1)$-dimensional space, associated with the above Lagrangian densities, expressed as functions of their respective variables, yields Eqs.~\eqref{eq:inhomFab} and \eqref{eq:inhomFa3}.

\subsection{Symmetries}
\label{firstdescentsym}

The invariance along the third spatial coordinate, that we assume in a specific inertial frame, is obviously not preserved by general Lorentz transformations. Lorentz transformations $\Lambda$ that maintain $z$-invariance in the transformed frame form the subgroup 
of matrices with the block-diagonal form
	\begin{align}\label{eq:firstdescentlambdafinal}
		\varLambda	=\left(\begin{BMAT}{c1c}{c1c}
			\vphantom{\begin{BMAT}[1pt]{c}{ccc} 0\\0\\0 \end{BMAT}}
			\mathmakebox[\widthof{$\begin{BMAT}[3pt]{ccc}{c}0&0&0\end{BMAT}$}]{\si L} & \\
			\vphantom{\cramped{(\si\eta\si b){}^\top }} & {\si Q}
		\end{BMAT}\right) ,
	\end{align}
with $\si L\in\LG[2]$ and $\si Q\in\Ort[1]=\{+1,-1\}$.

The matrices with ${\si Q} = 1$ are the elements of the isotropy group (little group) of the unit vector $(0,0,0,1)$ and in fact leave invariant all points of the $z$ axis. The matrices with ${\si Q}=-1$, instead, map $(0,0,0,1)$ to $(0,0,0,-1)$ and invert the $z$ axis. 

Observe that an inversion of the third axis implies that $\rho\to\rho$, $j^3\to-j^3$, $E^3\to-E^3$, $j^{a}\to j^{a}$, $E^{a}\to E^{a}$, for ${a}=1,2$. Moreover it can also be seen as the composition of the parity transformation, so that $\bm B\to\bm B$, and a rotation of $\pm\pi$ around the third axis, so that $B^{a}\to-B^{a}$, for $a=1,2$ whereas $B^3\to B^3$.
As a result, only a $(3+1)$-inertial observer can notice the effects of the sign of $\si Q$. Equations~\eqref{eq:homFa3}--\eqref{eq:inhomFa3} are unaffected by it, and each subsystem is manifestly invariant with respect to transformations in $2+1$ dimensions. We also understand the algebraic nature of the decompositions entailed by the block structures~\eqref{eq:tensordecomposition}, which correspond to the $3\oplus 3$ decomposition of the antisymmetric tensor field $F$ and the $3\oplus 1$ decomposition of the vector field $J$ with respect to $\LR[2]$ \cite{Hall,SU}.
	
\section{Second descent: from 2 to 1 spatial dimensions}
\label{seconddesc}

\subsection{Fields }
	\label{seconddescent}	
\begin{figure}
		\centering
		\begin{subfigure}[b]{0.48\textwidth} 
			\includegraphics[width=\linewidth]{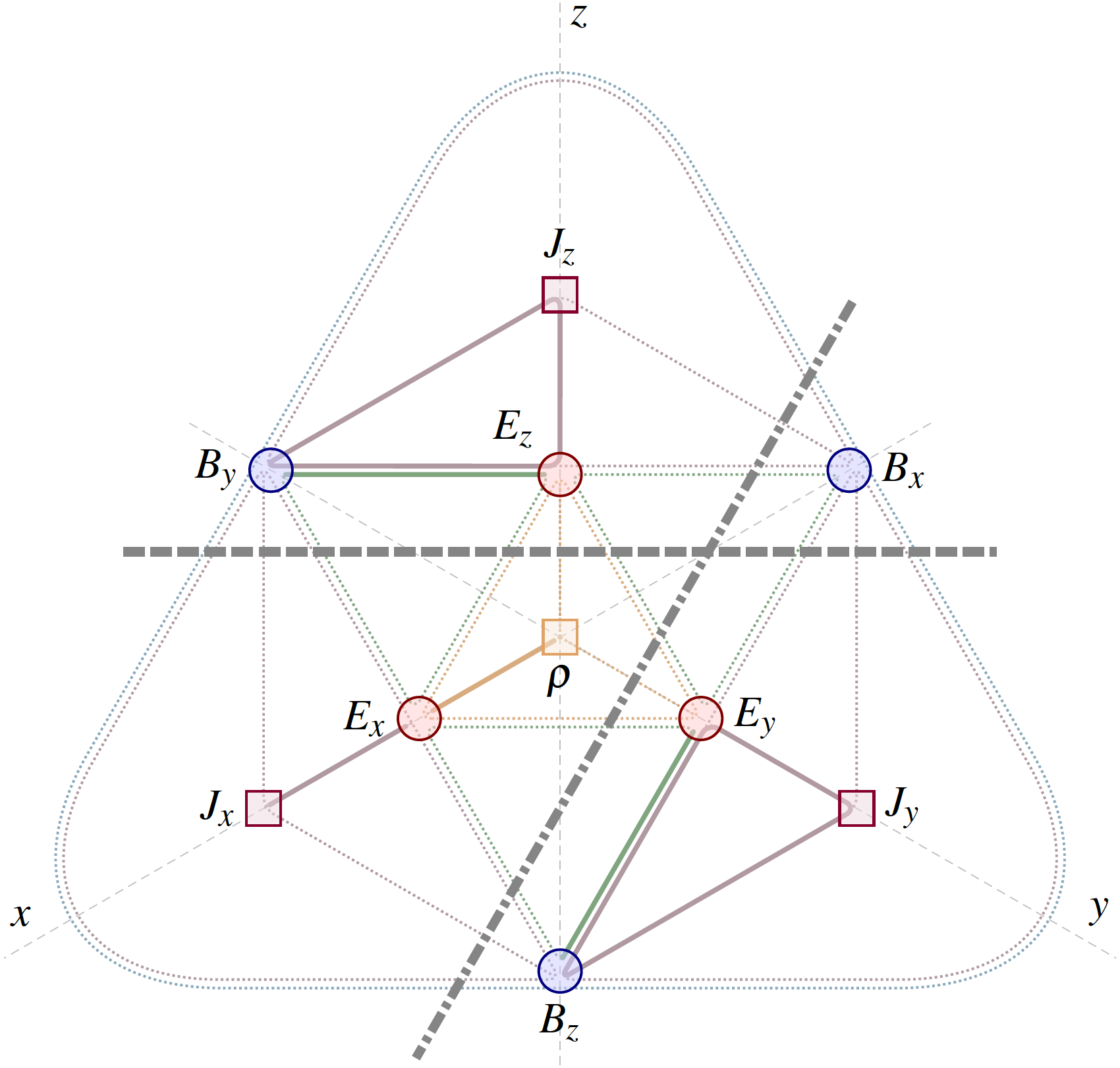}
			\caption{} \label{fig:secondadiscesa}
		\end{subfigure}\hfill
		\begin{subfigure}[b]{0.2211\textwidth} 
			\centering
			\includegraphics[width=\linewidth]{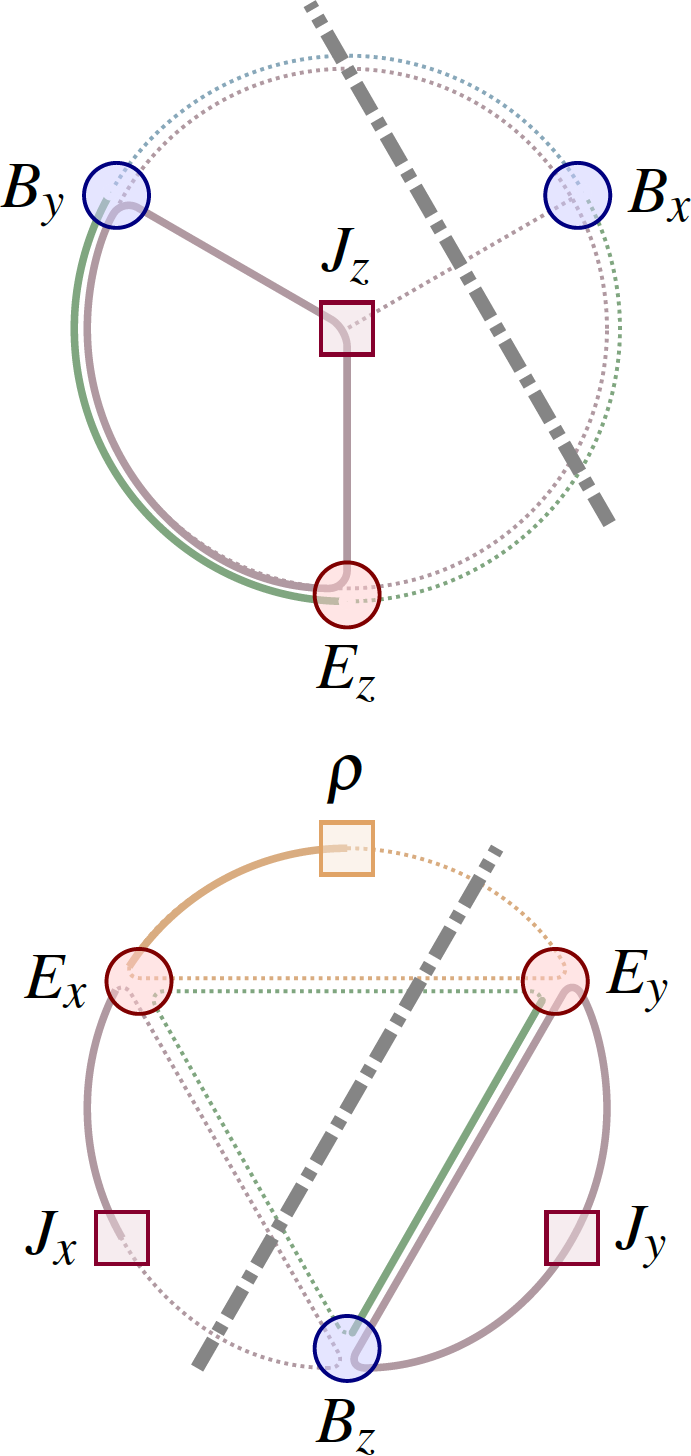}
			\caption{} \label{fig:secondadiscesaschema1}
		\end{subfigure}
		\hfill
		\begin{subfigure}[b]{0.2211\textwidth} 
			\centering
			\includegraphics[width=\linewidth]{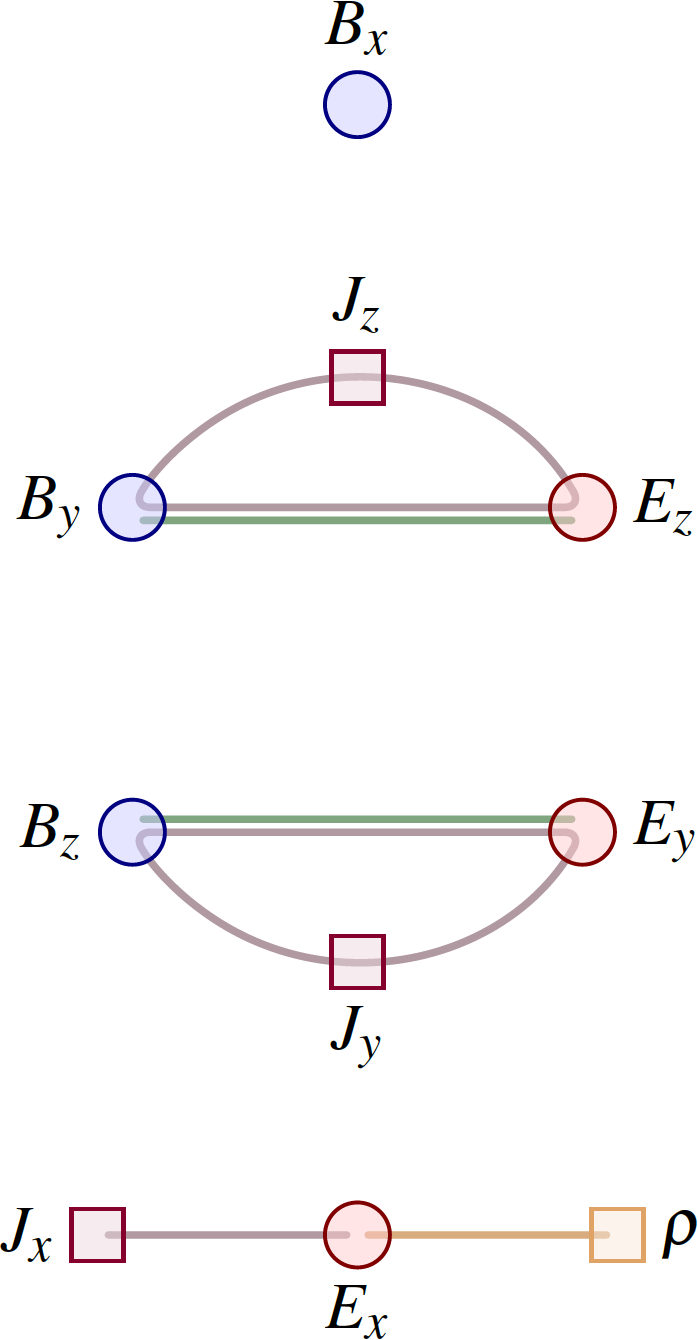}
			\caption{} \label{fig:secondadiscesaschema2}
		\end{subfigure}
		\caption{\raggedright Graph representation of the equations of the second descent. (a) Effects of the first (dashed line) and second (dash-dotted line) descent on the Maxwell graph. (b) Effects of the second descent on the graphs of the first descent. (c) Graphs of the equations of the second descent. In particular, {\sc magnetic Gauss}~(\protect\gaussmag) is reduced to the vertex $B_x$, and {\sc electric Gauss}~(\protect\gaussel) to the link between $E_x$ and $\rho$; each component of {\sc Faraday}~(\protect\faraday) is reduced to a single edge; as to {\sc Amp\`{e}re-Maxwell}~(\protect\ampere),  the $x$ component is reduced to the link between $E_x$ and $J_x$, while the other two components are reduced to the only nontrivial (triangular) subgraphs.}
		\label{fig:secondadiscesaglobale}
	\end{figure}

We now perform the second descent along the $y$ direction, by assuming that both the sources and the solutions of the equations of the first descent be also $y$-independent. Each subsystem,~\eqref{eq:ExEyBzi}--\eqref{eq:ExEyBzf} and~\eqref{eq:BxByEzi}--\eqref{eq:BxByEzf}, arising from the first descent, splits up into two uncoupled sets of equations, as graphically represented in Fig.~\ref{fig:secondadiscesaglobale}. Specifically, the sector $(E_x,E_y,B_z)$ is decoupled into $(E_x)$ and $(E_y,B_z)$, while the sector $(B_x,B_y,E_z)$ splits into $(B_y,E_z)$ and $(B_x)$.

\subsubsection{$(E_x)$ sector}		
\begin{figure}[h!t]
\centering
\begin{subfigure}[b]{.495\textwidth}
\includegraphics[width=\textwidth]{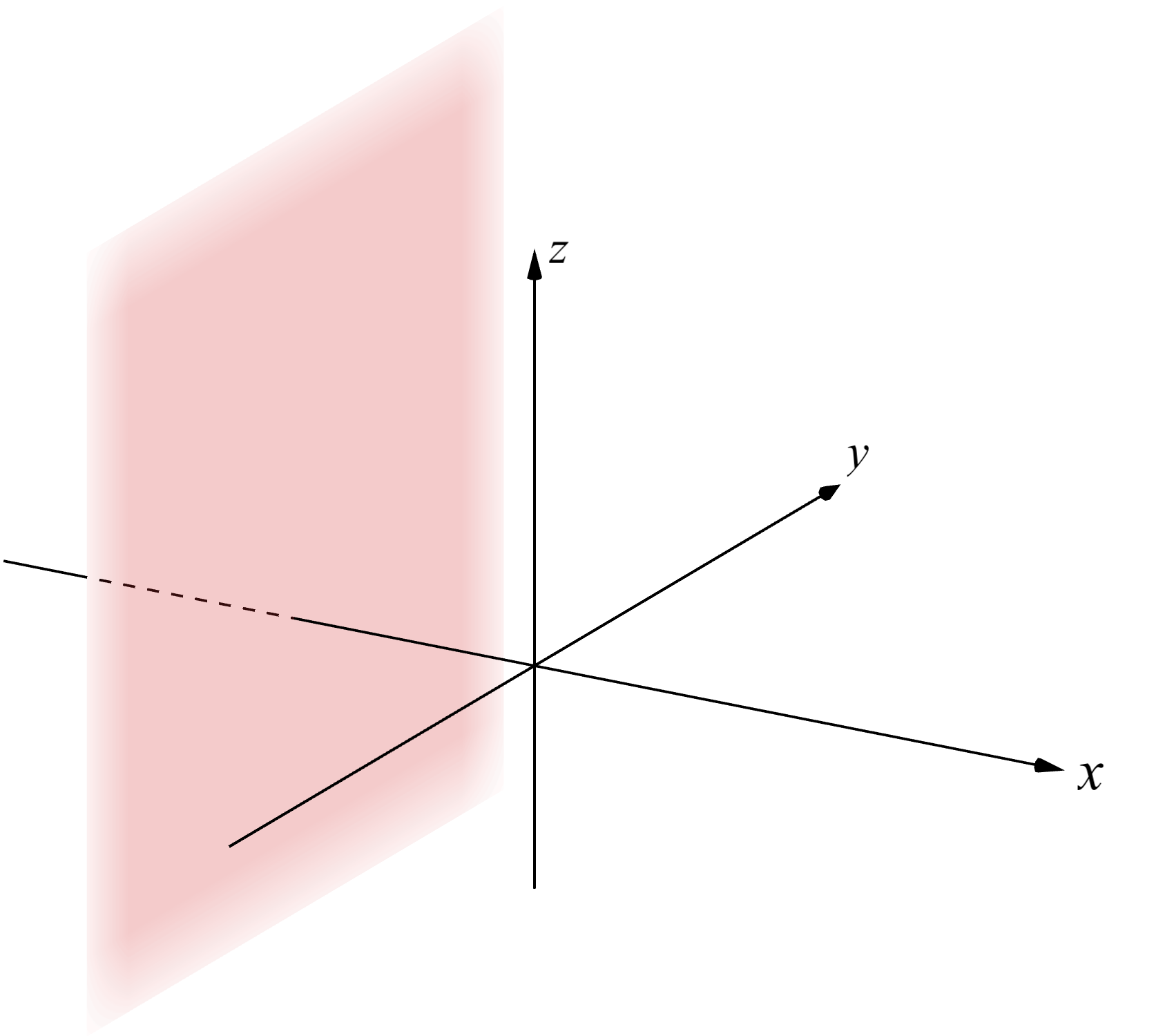}	
\caption{} \label{fig:3Dto1D_a}
\includegraphics[width=\textwidth]{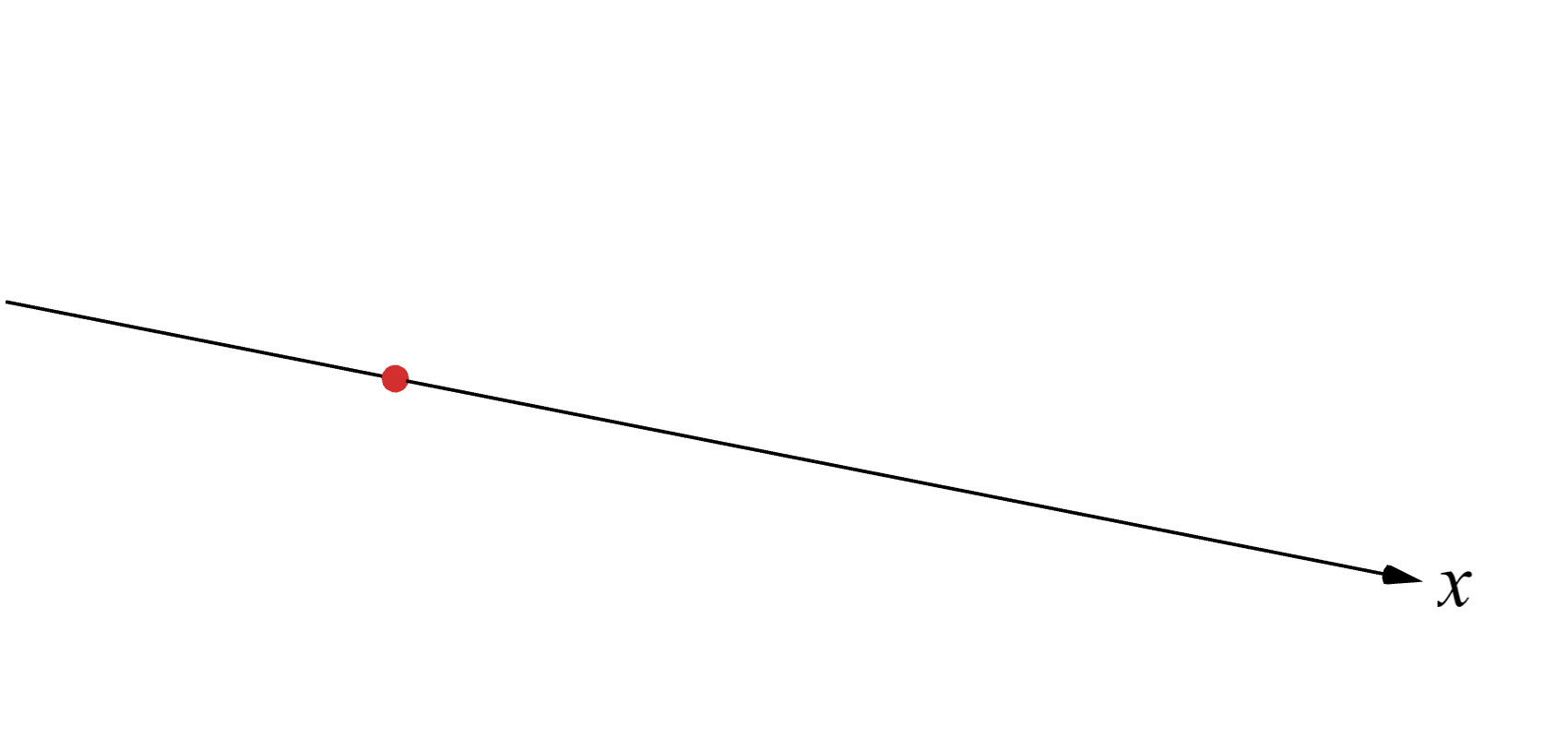}	
\caption{} \label{fig:3Dto1D_b}
\end{subfigure}	\hfill
\begin{subfigure}[b]{.495\textwidth}
\includegraphics[width=\textwidth]{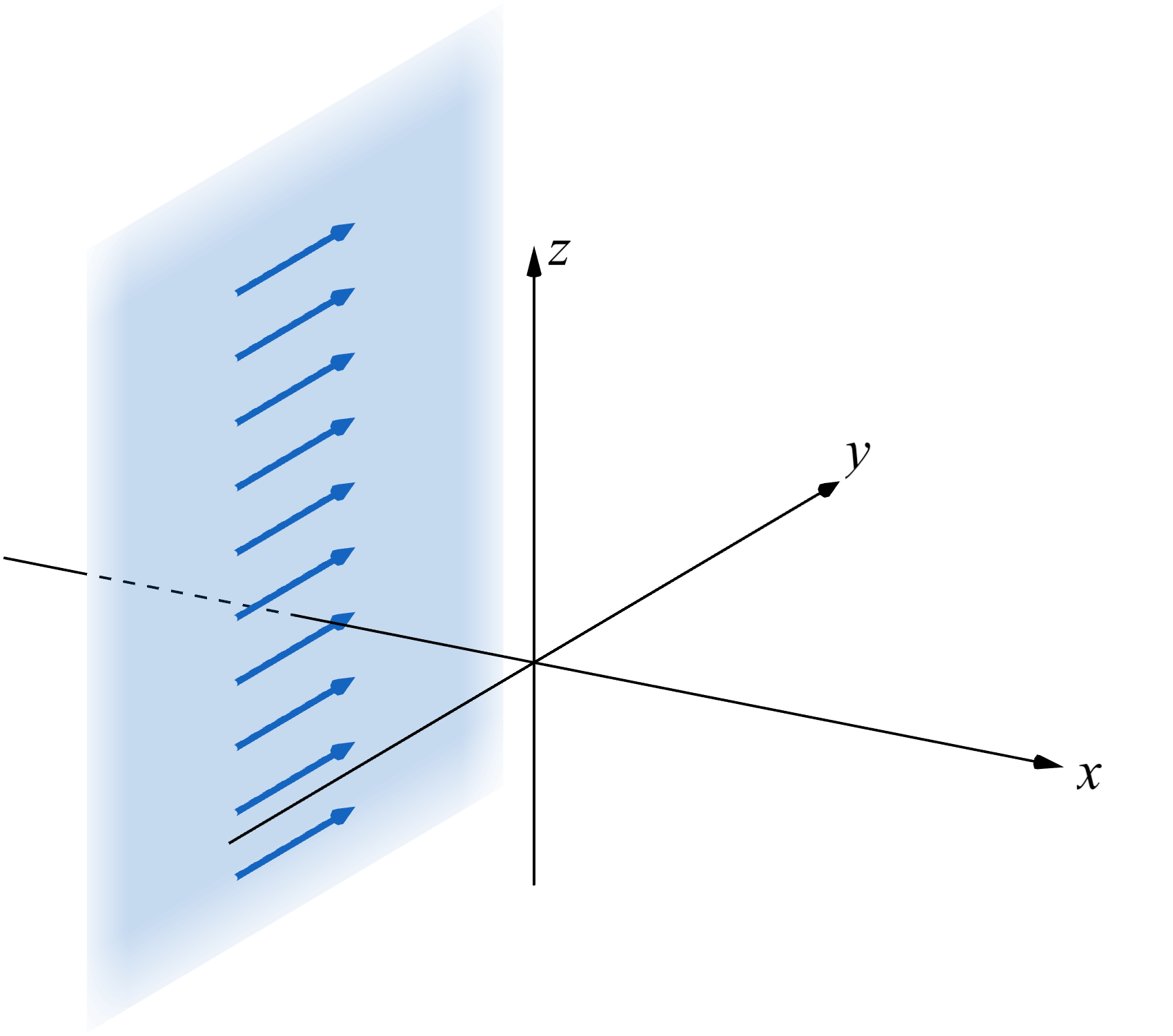}	
\caption{} \label{fig:3Dto1D_c}
\includegraphics[width=\textwidth]{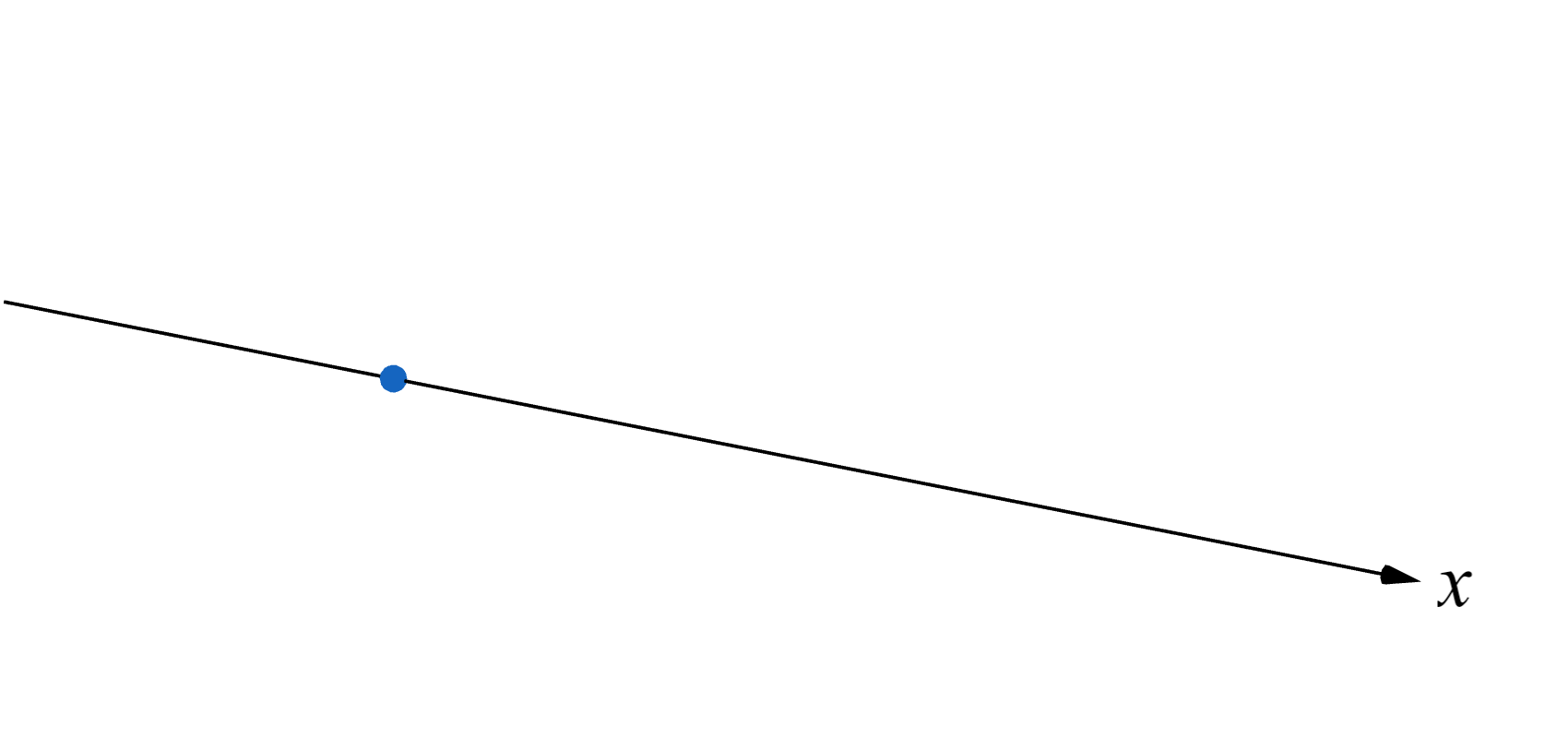}	
\caption{} \label{fig:3Dto1D_d}
\end{subfigure}	
\caption{\raggedright Top panels: $(y,z)$--independent sources in a three-dimensional space. Lower panels: their representation in $(1+1)$ electromagnetism. 
$(E_x)$ sector in $(1+1)$ electromagnetism: (a) the (red) uniform charged plane (b) is represented in one dimensions by a (red) point charge. 
$(E_y,B_z)$ sector in $(1+1)$ electromagnetism: (c) 
 the (blue) current plane, with current vector parallel to $z$, (d) becomes a (blue) point source. }
\label{fig:3Dto1D}
\end{figure}

The first sector is made of a single component $E_x$, governed by a system of two inhomogeneous equations
\begin{align}
	\partial_x{E_x}&= \rho,																										\label{eq:Exi}\\
	-\partial_t{E_x}&= J_x .																										\label{eq:Exf}
\end{align}
The source terms $\rho$ and $J_x$ are the only components of the source vector that are still linked by the remnants 
\begin{equation}
    \partial_t \rho+\partial_xJ_x=0
\end{equation}
of the $(3+1)$-dimensional continuity equation, that can be also obtained by direct derivation of Eqs.~\eqref{eq:Exi}--\eqref{eq:Exf}. Notice that the  dynamics in the $(E_x)$ sector does not depend on the speed of light, that has canceled out in Eq.~\eqref{eq:Exf}. The general solution 
\begin{equation}
    E_x(t,x) = E_x(t_0,x_0) + \int_{x_0}^x dx' \rho(t,x') - \int_{t_0}^t dt' J_x(t',x_0)
\end{equation}
confirms that no wave propagation occurs in this sector~\cite{McDonald}. A pictorial representation of the sources is given in Fig.\ \ref{fig:3Dto1D}(a)-(b).
This is the model that is normally adopted in the quantization of EM in 1+1 dimensions \cite{hist1,hist2,hist3}.

\subsubsection{$(E_y, B_z)$ sector}		
The components $(E_y,B_z)$ satisfy a system of one homogeneous and one inhomogeneous equation,
\begin{align}
	\frac{1}{c}\partial_t{B_z}+\partial_x{E_y}&=0,		\label{eq:EyBzi}\\
	-\frac{1}{c}\partial_t{E_y}-\partial_x{B_z}&=\frac{1}{c}J_y ,		\label{eq:EyBzf}
\end{align}
with source term $J_y$. The solutions of the above equations correspond to transverse and linearly polarized EM waves propagating along the $x$ axis,
\begin{align}
	\frac{1}{c^2}\partial_t^2{E_y}-\partial_x^2{E_y}&=-\frac{1}{c^2} \partial_t J_y,		\\
	\frac{1}{c^2}\partial_t^2{B_z}-\partial_x^2{B_z}&=\frac{1}{c}\partial_x J_y.		
\end{align}
This theory, admitting wave propagation, is not the model normally adopted to discuss the quantization of EM in 1+1 dimensions. Notice also that this theory involves an equal number of electric and magnetic components, a feature that, under Ehrenfest's assumption\cite{Ehrenfest}, would be unique to the $3+1$ case. A pictorial representation of the sources is given in Fig.\ \ref{fig:3Dto1D}(c)-(d).

\subsubsection{$(B_y, E_z)$ sector}		
The components $(B_y,E_z)$, obey a similar pair of equations to those in the $(E_y,B_z)$ sector,
\begin{align}
	\frac{1}{c}\partial_t{B_y}-\partial_x{E_z}&=0,	\label{eq:ByEzi}\\
	-\frac{1}{c}\partial_t{E_z}+\partial_x{B_y}&=\frac{1}{c}J_z . 	\label{eq:ByEzf}
\end{align}
with source term $J_z$. Also in this case, the free solutions are linearly polarized waves,
\begin{align}
	\frac{1}{c^2}\partial_t^2{E_z}-\partial_x^2{E_z}&=-\frac{1}{c^2} \partial_t J_z,		\\
	\frac{1}{c^2}\partial_t^2{B_y}-\partial_x^2{B_y}&=-\frac{1}{c}\partial_x J_z.		
\end{align}
Sectors  $(E_y,B_z)$ and $(B_y,E_z)$ are connected by a $\pi/2$ rotation of fields and source around the $x$ axis, and thus they are described by the same EM theory.

\subsubsection{$(B_x)$ sector}		

The single component $B_x$ satisfies a system of two homogeneous equations
\begin{align}
	\partial_x{B_x}&=0, \label{eq:Bxi} \\
	\partial_t{B_x}&=0,	\label{eq:Bxf}
\end{align}
that make the sector trivial, since the only solution is a uniform and constant magnetic field. Notice that $B_x$ is completely fixed by non-dynamical equations and by the condition of $(y,z)$-independence.

\subsection{Tensor notation}\label{2descenttensor}		
After the second descent, each block of the first descent splits up into two sub-blocks. Specifically, 
fields and sources separate as follows:
	\begin{align}\label{eq:secondtensordecomposition}
		(F^{\mu\nu})=\left(
		\begin{BMAT}{cc1cc}{cc1cc}
			0 & \textcolor{red!50!black}{-E_x} & \textcolor{orange!75!black}{-E_y} & \textcolor{blue!50!black}{-E_z}\\
			\textcolor{red!50!black}{E_x} & 0 & \textcolor{orange!75!black}{-B_z} & \textcolor{blue!50!black}{B_y}\\
			\textcolor{orange!75!black}{E_y} & \textcolor{orange!75!black}{B_z} & 0 &\textcolor{cyan!60!black}{-B_x}\\
			\textcolor{blue!50!black}{E_z} & \textcolor{blue!50!black}{-B_y} & \textcolor{cyan!60!black}{B_x} & 0
			\addpath{(0,1,1)rr}
			\addpath{(3,4,1)dd}
		\end{BMAT}\right),&&
		(G^{\mu\nu})=\left(
		\begin{BMAT}{cc1cc}{cc1cc}
			0 & \textcolor{cyan!60!black}{-B_x} & \textcolor{blue!50!black}{-B_y} & \textcolor{orange!75!black}{-B_z}\\
			\textcolor{cyan!60!black}{B_x} & 0 & \textcolor{blue!50!black}{E_z} & \textcolor{orange!75!black}{-E_y}\\
			\textcolor{blue!50!black}{B_y} & \textcolor{blue!50!black}{-E_z} & 0 &\textcolor{red!50!black}{E_x}\\
			\textcolor{orange!75!black}{B_z} & \textcolor{orange!75!black}{E_y} & \textcolor{red!50!black}{-E_x} & 0
			\addpath{(0,1,1)rr}
			\addpath{(3,4,1)dd}
		\end{BMAT}\right),
		&&
		(j^{\mu})=\left(\begin{BMAT}{c}{cc1c1c}
			\textcolor{red!50!black}{c\rho}\\
			\textcolor{red!50!black}{J_x}\\
			\textcolor{orange!75!black}{J_y}\\
			\textcolor{blue!50!black}{J_z}
		\end{BMAT}\right).
	\end{align}
The elements of each sector pair $[(E_x),(B_x)]$ and $[(E_y,B_z),(B_y,E_z)]$ are related to each other by duality: this result follows from the interplay between $(3+1)$- and $(1+1)$-duality for a rank-2 antisymmetric $(3+1)$-tensor, yielding
\begin{align}\label{eq:secondduality}
	G^{{rs}}
	&=\frac{1}{2}\qty(\epsilon^{{rs} 23 }F_{23}+\epsilon^{{rs}32}F_{32})
	=\epsilon^{{rs}}F_{23}=-\epsilon^{{rs}} B_x,\\
	G^{{r}2}
	&=\frac{1}{2}\qty(\epsilon^{{r} 2 {s} 3}F_{{s}3}+\epsilon^{{r}2 3{s}}F_{3{s}})
	=-\epsilon^{{rs}}F_{{s} 3},\\
	G^{{r} 3}
	&=\frac{1}{2}\qty(\epsilon^{{r}3{s}2}F_{{s}2}+\epsilon^{{r} 32{s}}F_{2{s}})
	=\epsilon^{{rs}}F_{{s}2},\\
	G^{2 3}&=\frac{1}{2}\epsilon^{23{rs}}F_{{rs}}=F_{01}=E_x,\label{eq: 3+1 vs 1+1 duality 4}
\end{align}
where henceforth the latin indices $r,s$ take values in $\{0,1\}$. 

In tensor form, the equations for the $(E_x)$ sector reads
\begin{align}\label{eq:electric}
	\partial_{{r}} F^{{rs}}&=\frac{1}{c} j^{{s}},
\end{align}
The sector involves a tensor field in $(1+1)$ dimensions 
\begin{equation}\label{11tensor}
(F^{{rs}}) =
\begin{pmatrix}
   0       & -E_x  \\
  E_x     & 0 \\
\end{pmatrix} ,
\end{equation}
its scalar dual, $G^{23}=E_x$, and the $(1+1)$-vector source $(j^{{r}})=(c\rho,J_x)$. The $(E_y,B_z)$ sector is characterized by the tensor equations
\begin{align}
	\epsilon^{{rs}}\partial_{{r}} F_{{s}2}&=0,										\label{eq:mixed1i}\\
	\partial_{{r}} F^{{r}2}&=\frac{1}{c} j^2,										  \label{eq:mixed1f}
\end{align}
equivalent to the system~\eqref{eq:EyBzi}--\eqref{eq:EyBzf}. In this case, both the field $(F^{{r}2})=(-E_y,-B_z)$ and its dual $(-G^{{r}3})=(B_z,E_y)$ are vectors, while the source $j^2=J_y$ behaves as a scalar. This sector is related by duality to $(B_y,E_z)$, characterized by the equations
\begin{align}
	\epsilon^{{rs}}\partial_{{r}} F_{{s}3}&=0,									\label{eq:mixed2i}\\
	\partial_{{r}} F^{{r}3}&=\frac{1}{c} j^3,										\label{eq:mixed2f}
\end{align}
and characterized as well by a vector field $(F^{{r}3})=(-E_z,B_y)$, a vector dual 
$(G^{{r}2})=(-B_y,E_z)$, and a scalar source $j^3=J_z$. Finally, the tensor form of the equations for the $B_x$ sector reads
\begin{align}
	\epsilon^{{rs}}\partial_{{r}} F_{23}=0,													\label{eq:magnetic}
\end{align}
where the scalar field $F^{23}=-B_x$ corresponds to the $(1+1)$ tensor dual 
\begin{align}\label{11tensor}
(G^{{rs}}) & =
\begin{pmatrix}
   0       & -B_x  \\
  B_x     & 0 \\
\end{pmatrix} .
\end{align}
The non-dynamical nature of this last sector is confirmed by the absence of sources. The fields in all sectors formally obey a wave equations in $(1+1)$ dimensions, characterized by the d'A\-lem\-bert operator $\partial_{{u}}\partial^{{u}} = c^{-2}\partial_t^2 - \partial_x^2$ 
\begin{align}
\partial_{{u}}\partial^{{u}} F^{{rs}} &=
	\frac{1}{c}(\partial^{{r}}  j^{{s}}-\partial^{{s}}  j^{{r}}),					\label{eq:secondwaveselectric}\\
\partial_{{u}}\partial^{{u}} F^{{r}2} &=\frac{1}{c}\partial^{{r}} j^2,\\
\partial_{{u}}\partial^{{u}} F^{{r}3} &=\frac{1}{c}\partial^{{r}} j^3,\\
\partial_{{u}}\partial^{{u}} F^{23} &=0,					\label{eq:secondwavesmagnetic}
\end{align}
However, as we already remarked, wave propagation actually occurs only in the sectors in which two fields are featured, while in the $(E_x)$ and $(B_x)$ sectors the above wave equations are redundant. 

Remarkably, the two sectors that involve both electric and magnetic field, formally equivalent to each other, arise from different first descent's subsystem. They also have several features in common with the general EM theory in $(3+1)$ dimensions: for instance, i) the number of electric and magnetic components is the same; ii) the number of homogeneous and inhomogeneous equations is the same; iii) the tensor $F$ and its dual $G$ are mathematical objects of the same kind; iv) there is wave propagation. On the other hand, the mathematical objects of the two remaining subsystems are highly unbalanced, as $(E_x)$ is characterized by a tensor field, a scalar dual, and no homogeneous equation, while $(B_x)$ has a scalar field, a tensor dual, and no inhomogeneous equation.

\subsection{Potentials and Lagrangian}
\label{potl1}

In view of the second descent, the Lagrangian~\eqref{eq:L3D} can be conveniently decomposed as follows:
\begin{align}
 \mathcal{L}_{3+1} 
	= & \left(-\frac{1}{4}F_{{rs}}F^{{rs}}-\frac{1}{c}j^{r}A_{r}\right)
	+ \left(-\frac{1}{2}F_{{r2}}F^{{r2}}-\frac{1}{c}j^{2}A_{2}\right) \nonumber \\
	& +  \left(-\frac{1}{2}F_{{r3}}F^{{r3}}-\frac{1}{c}j^{3}A_{3}\right) 
	+ \left( -\frac{1}{2}F_{{23}}F^{{23}} \right) \\
	=  & \; \mathcal L_{E} + \mathcal L_{EB} + \mathcal L_{BE} +  \mathcal L_{B} .
	 \label{eq:Ldec1d}
\end{align}
The dynamics in the $(E_x)$, $(E_y,B_z)$, and $(B_y,E_z)$ sectors can be obtained from the Euler-Lagrange equations determined by the terms $\mathcal{L}_E$, $\mathcal{L}_{EB}$ and $\mathcal{L}_{BE}$, respectively, combined with the assumption of field independence from $y$ and $z$. The term $\mathcal{L}_B=-B_x^2/2$, instead, is fixed to a constant by the homogeneous Maxwell's equations. 
 The four parts of the Lagrangian, however, are not decoupled as functions of the potentials, since $A_0$ and $A_1$ appear in both $\mathcal{L}_{EB}$ and $\mathcal{L}_{BE}$ through their derivatives $\partial_2A_{{r}}$ and $\partial_3A_{{r}}$, respectively. As in the case of the dimensional reduction to $(2+1)$, we can choose the gauge fixing  $\partial_2A_{{r}}=\partial_3A_{{r}}=0$, which decouples the four sectors. This leaves the gauge freedom $A_{{r}}\to A_{{r}}+\partial_{{r}}f$, with $f=f(t,x)$.  

In the decoupled case, the Euler-Lagrange equations in $(1+1)$ dimensions associated to the Lagrangian density
\begin{equation}\label{eq:lage}
	\mathcal{L}_{E}(A^0,A^1) = -\frac{1}{4}(\partial_{{r}}A_{{r}} - \partial_{{s}}A_{{r}})(\partial^{{r}}A^{{s}} - \partial^{{s}}A^{{r}}) -\frac{1}{c}j^{r}A_{r} 
	= \frac{1}{2} E_x^2- \rho \Phi +\tfrac{1}{c} J_x A_x
\end{equation}
yield the dynamics~(\ref{eq:Exi})--(\ref{eq:Exf}) in the $(E_x)$ sector, while the second term 
\begin{equation}
\label{eq:lageb}
	\mathcal L_{EB} (A^2) = -\frac{1}{2}( \partial_{{r}}A_2 ) ( \partial^{{r}}A^2 ) -\frac{1}{c}j^2 A_2
	=\frac{1}{2}\qty(E_y^2- B_z^2)+\frac{1}{c} J_y A_y 
\end{equation}
and the third term
\begin{equation}
\label{eq:lagbe}
	\mathcal L_{BE} (A^3) = -\frac{1}{2}( \partial_{{r}}A_3 ) ( \partial^{{r}}A^3 ) -\frac{1}{c} j^3 A_3
	= \frac{1}{2}\qty(E_z^2- B_y^2)+\frac{1}{c} J_z A_z 
\end{equation}
give the dynamical equations \eqref{eq:EyBzf} and \eqref{eq:ByEzf} of the transverse fields in the $(E_y,B_z)$ and $(B_y,E_z)$ sectors, respectively.

	\subsection{Symmetries}

After the second descent, a Lorentz transformation takes the block-diagonal form
	\begin{align}\label{eq:secondlambda}
		\varLambda	=\left(\begin{BMAT}[3pt]{c1c}{c1c}
			\vphantom{\begin{BMAT}[1pt]{c}{cc} 0\\0 \end{BMAT}}
			\mathmakebox[\widthof{$\begin{BMAT}[3pt]{cc}{c}0&0\end{BMAT}$}]{\si L} & \\
			\vphantom{\cramped{(\si\eta\si b){}^\top }} & 
			\vphantom{\begin{BMAT}[1pt]{c}{cc} 0\\0 \end{BMAT}}
			\mathmakebox[\widthof{$\begin{BMAT}[3pt]{cc}{c}0&0\end{BMAT}$}]{\si Q}
		\end{BMAT}\right),
	\end{align}
with $\si L\in\LG[1]$, $\si Q\in\Ort[2]$. 
The matrices $\Lambda$ form the isotropy subgroup of the $yz$-plane, that is such that $\Lambda (0,0,y,z)^{\top}=(0,0,y',z')^{\top}$.

We also realize the algebraic nature of the decompositions entailed by the block-structures after the first, Eq.~\eqref{eq:tensordecomposition}, and the second, Eq.~\eqref{eq:secondtensordecomposition}, reduction. The former are precisely the $3\oplus 3$ decomposition of the antisymmetric tensor $(F^{\mu\nu})$ and the $3\oplus 1$ decomposition of the vector $(j^\mu)$ with respect to $\LG[2]$, see the decomposition~\eqref{eq:firstdescentlambdafinal}. 
The latter are the $2\oplus 2\oplus 1\oplus 1$ decomposition of the antisymmetric tensor $F$ and the $2\oplus 1\oplus 1$ decomposition of the vector $j$ with respect to $\LG[1]$, according to the splitting~\eqref{eq:secondlambda}.

\section{Conclusions and outlook}
\label{concl}

By applying  dimensional reduction, we found that classical electromagnetism in $(3+1)$ dimensions splits into two different theories in $(2+1)$ dimensions, if one assumes invariance of fields and sources with respect to one space direction. A further reduction, obtained by assuming invariance with respect to a plane, yielded four independent sectors described by three different EM theories in $(1+1)$ dimensions. 
 One of the theories in each reduction, namely $(E_x,E_y,B_z)$ in $(2+1)$ dimensions and $(E_x)$ in $(1+1)$ dimensions, appears in low-dimensional generalizations of Maxwell's equations~\cite{hist1}, and actually their dynamics is governed by tensor equations of the same form as those in $(3+1)$ dimensions. However, these models do not predict the remaining EM theories, that are a consequence of the fact that low-dimensional theories are still embedded in a $(3+1)$-dimensional spacetime. Future research will be dedicated to a deeper analysis of the gauge-invariance breaking patterns in the reduction, and to the role of parity symmetry (which would require the use of the $D$ and $H$ fields \cite{Sommerfeld}).
In addition, we plan to investigate the role of dimensional reduction in non-Abelian theories and systems in which (classical and quantum) fields interact with dynamical sources.

\begin{acknowledgments}
This research was funded by MIUR via PRIN 2017 (Progetto di Ricerca di Interesse Nazionale), project QUSHIP (2017SRNBRK), by the Italian National Group of Mathematical Physics (GNFM-INdAM), by Istituto Nazionale di Fisica Nucleare (INFN) through the project ``QUANTUM'', and by Regione Puglia and QuantERA ERA-NET Cofund in Quantum Technologies (Grant No.\ 731473), project QuantHEP\@.

\end{acknowledgments}

%
%

\nocite{*}

\end{document}